\documentclass[conference]{IEEEtran}
\usepackage[latin9]{inputenc}
\usepackage{amsmath}
\usepackage{graphicx}

\usepackage{euscript}
\usepackage{amsmath}
\usepackage{amsthm}
\usepackage{amssymb}
\usepackage{booktabs}
\usepackage{epsfig}
\usepackage{xspace}
\usepackage{graphicx}
\usepackage{epstopdf}
\usepackage{epsfig}
\usepackage{color}
\usepackage{url}
\graphicspath{{Images/}}
\usepackage{multirow}
\usepackage{cite}
\usepackage[numbers,sort&compress]{natbib}
\usepackage[table,xcdraw]{xcolor}
\usepackage{overpic}

\usepackage{enumitem,kantlipsum}
\usepackage[ruled,vlined,linesnumbered]{algorithm2e}

\usepackage{enumitem}
\setlist[enumerate]{label*=.\arabic*, after=\normalfont}
\setlist[enumerate,1]{label=\arabic*, font=\normalfont\color{red}}
\setlist[enumerate,2]{before=\normalfont}
\setlist[enumerate,3]{font=\normalfont\upshape, before=\normalfont\itshape}

\usepackage{tikz}
\usepackage{collcell}

\ifCLASSINFOpdf
\else
\fi

\usepackage{stfloats}
\begin{document}
\title{Empirical Evaluation of PRNU Fingerprint Variation for Mismatched Imaging Pipelines}

% Yo
% \author{\IEEEauthorblockN{Sharad Joshi}
% \IEEEauthorblockA{MANAS Lab \\IIT Gandhinagar, India \\
% Email: sharad.joshi@iitgn.ac.in}
% \and
% \IEEEauthorblockN{Pawel Korus}
% \IEEEauthorblockA{Tandon School of Engineering\\
% New York University, USA\\
% Email:  pkorus@nyu.edu}
% \and
% \IEEEauthorblockN{Nitin Khanna}
% \IEEEauthorblockA{ MANAS Lab \\IIT Gandhinagar, India \\
% Email:  nitinkhanna@iitgn.ac.in}
% \and
% \IEEEauthorblockN{Nasir Memon}
% \IEEEauthorblockA{Tandon School of Engineering\\
% New York University, USA\\
% Email:  memon@nyu.edu}}

\author{\IEEEauthorblockN{Sharad Joshi$^{1}$, Pawel Korus$^{2,3}$, Nitin Khanna$^{1}$, and Nasir Memon$^{2}$}
\IEEEauthorblockA{$^{1}$MANAS Lab, Indian Institute of Technology Gandhinagar, India\\ $^{2}$Tandon School of Engineering, New York University, USA\\ $^{3}$Department of Telecommunications, AGH University of Science and Technology, Poland\\ e-mail: \{sharad.joshi,nitinkhanna\}@iitgn.ac.in, \{pkorus,memon\}@nyu.edu\\}
}

% \and
% \IEEEauthorblockN{Author 3 and Author 4}
% \IEEEauthorblockA{Affiliation\\
% Affiliation\\
% Email:  \{author3,author4\}@anon.myo.us}}


%Telephone: (800) 555--1212, Fax: (888) 555--1212}


\maketitle

%INCLUDES COPYRIGHT NOTICE: one of three copyright notice should be included.
%Uncomment the appropriate line below, according to the authors %affiliation:
% \begin{figure}[b]
% \vspace{-0.3cm}
% \parbox{\hsize}{\em
% %information about the event:
% WIFS`2020, December, 6-11, 2020, New York, USA.
% %copyright notice: one of four copyright notices below should be included. Choose the right one below according to the authors affiliation:
% %XXX-X-XXXX-XXXX-X/XX/\$XX.00 \ \copyright 2017 European Union.
% %XXX-X-XXXX-XXXX-X/XX/\$XX.00  \ \copyright 2017 Crown.
% %U.S. Government work not protected by U.S. copyright.
% %XXX-X-XXXX-XXXX-X/XX/\$XX.00 \ \copyright 2017 IEEE.
% 978-1-7281-9930-6/20/\$31.00 \ \copyright 2020 IEEE.
% }\end{figure}

\begin{abstract}
We assess the variability of PRNU-based camera fingerprints with mismatched imaging pipelines (e.g., different camera ISP or digital darkroom software). We show that camera fingerprints exhibit non-negligible variations in this setup, which may lead to unexpected degradation of detection statistics in real-world use-cases. 
We tested 13 different pipelines, including standard digital darkroom software and recent neural-networks. We observed that correlation between fingerprints from mismatched pipelines drops on average to 0.38 and the PCE detection statistic drops by over 40\%. The degradation in error rates is the strongest for small patches commonly used in photo manipulation detection, and when neural networks are used for photo development. At a fixed 0.5\% FPR setting, the TPR drops by 17 ppt (percentage points) for 128~px and 256~px patches.
\end{abstract}

% \fi
\IEEEpeerreviewmaketitle

\section{Introduction}

\IEEEPARstart{C}{amera} fingerprints based on photo-response non-uniformity (PRNU) of the imaging sensor are one of the most reliable tools in photo forensics~\cite{MemonBook2013,piva2013overview,korus2017digital}. They have been studied in the literature for over a decade~\cite{lukas2006digital,kang2014context,zhang2017identifying,al2017spn} and are commonly used in two main applications: (1) \emph{source attribution} - which links photographs to a specific instance of a camera; and (2) \emph{photo manipulation detection} - which involves a local analysis of the sensor fingerprint to identify areas of mismatch~\cite{chen2008determining}.

The success of PRNU fingerprints stems from their stability and robustness to various post-processing that happens in real-world applications (e.g., scaling, or lossy compression). Some of this post-processing, in particular the impact of JPEG compression or denoising, have been extensively studied~\cite{quiring2015fragile,quiring2019security,cortiana2011performance,chierchia2010influence}. However, many others remain unexplored. This may lead to inaccurate detection and under-estimated error rates.

In this work, we empirically assess the sensitivity of PRNU fingerprints to \emph{mismatched imaging pipelines}. This scenario arises when test images are processed differently (e.g., using 3rd-party darkroom software like LibRAW or Lightroom) than the images used for fingerprint estimation. We believe this scenario will become increasingly more common with the progress of machine learning and the increasing adoption of deep neural networks in various stages of the imaging pipeline (e.g., demosaicing~\cite{Gharbi2016}, denoising~\cite{lehtinen2018noise2noise}, or tone-mapping~\cite{Eilertsen2017}) or even instead of it~\cite{ignatov2020replacing}. Moreover, modern darkrooms, e.g., \emph{Luminar}, extensively advertise ML-based solutions in their image processing and enhancement routines, and such a trend is likely to increase.

\begin{figure}[t]
        \begin{center}
        \includegraphics[width=\linewidth]{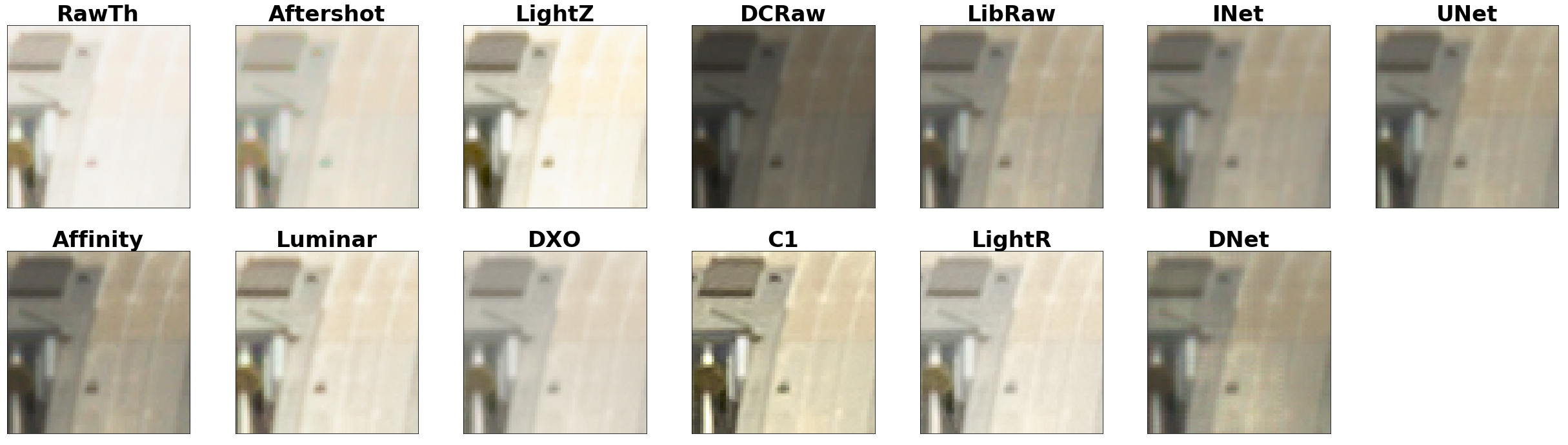}
        \includegraphics[width=\linewidth]{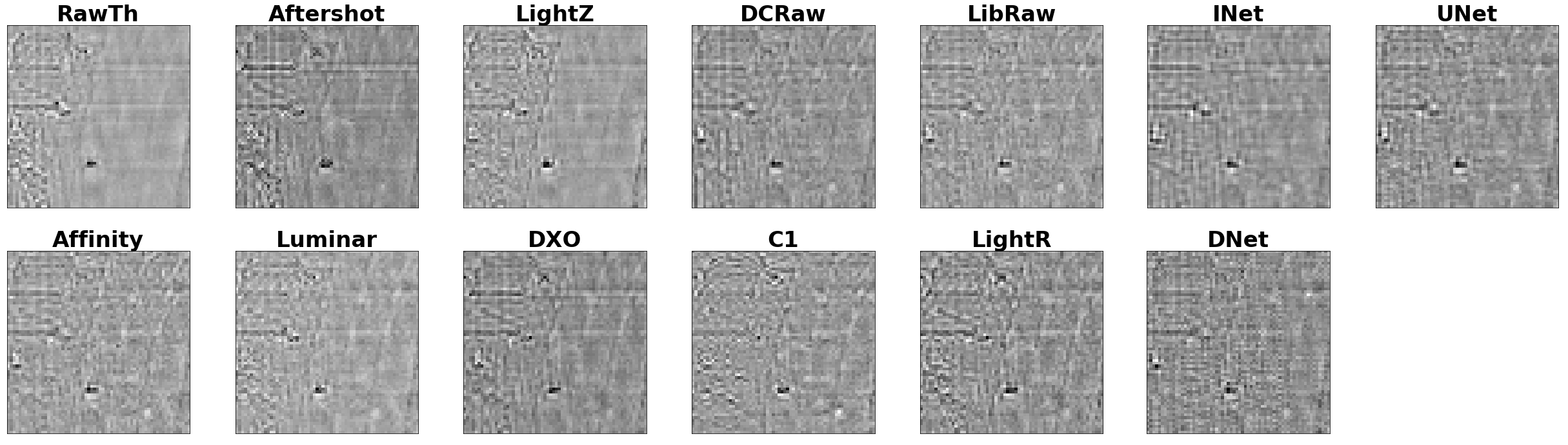}
    \caption{Image patches of size 64 $\times$ 64 px (top) and their residuals (bottom) from 13 different imaging pipelines operating with default settings.}
    \label{fig:InputImages}
        \end{center}
\end{figure}

In this work, we take the first step in this direction and assess the impact of image signal processor (ISP) i.e., imaging pipeline  variation on PRNU fingerprint analysis. We used 13 software pipelines representing popular digital darkrooms and 3 neural network architectures. We consider a scenario where each pipeline uses default settings with minimal post-processing, and neural-network is trained to reproduce images visually equivalent to a standard camera pipeline~\cite{korus2019neural,korus2019content} (see Figure~\ref{fig:InputImages} for example image patches and residuals for Nikon D7000).

The main observations from our work are as follows:
\begin{itemize}
    \item{Correlation between camera fingerprints obtained from exactly the same images processed by different pipelines drops on average by 62 \%.}
    \item{Degradation in detection statistics is the strongest for smaller image patches (128 or 256~px windows), which will be particularly detrimental to manipulation detection.}
    \item{For large patches (1024~px), standard ISPs reveal little variation in detection statistics, but the effect remains strong for neural ISPs.}
    \item{For 128~px windows commonly used in manipulation detection, we observed an average deterioration of 41\% in median PCE and of 17\% in TPR at 0.5\% FPR.}
    \item{Imaging pipelines based on neural networks tend to distort the camera fingerprints more - TPR dropped by 20\% compared to 14\% for standard pipelines.} 
    \item{Even for traditional pipelines, we observed unexpected matching failures for some configurations (\emph{RawTherapee} and \emph{Capture One} software with Nikon cameras).}
\end{itemize}

\section{Background and Related Work}
\label{sec:PRNU}

% What PRNU is and how it works
% Typically estimated in RGB / often only green channel used    
% or only measured (not interpolated colors) \cite{li2011color}

A PRNU fingerprint of a camera characterizes the consistent bias of individual pixels in its imaging sensor~\cite{chen2008determining}. The fingerprint is estimated from multiple images by carefully averaging high-frequency residuals where a denoising filter suppresses content. 
At first, a noise residual $R_i$ is computed for each input image $I_{i}$ using a denoising filter $D$ as $R_{i} = I_{i} - D(I_{i}), i=1,..,N.$
In the next step, the fingerprint $k$ is estimated using maximum likelihood estimation~\cite{chen2008determining}:
\begin{equation}
\label{eq:ISP_eq2}
 k = \frac{\sum\limits_{i=1}^{N}R_{i}I_{i}}{\sum\limits_{i=1}^{N}(I_{i})^2} \\
\end{equation}
At test time, the corresponding residual of a test image is correlated with the camera fingerprint - either globally (for attribution) or locally (for manipulation detection).

% {\color{red}We need more about other lines of work - PRNU enhancement, better denoisers, CNNs, correlation predictors... If there is space, feel free to put some basic PRNU equations.}
Several enhancement techniques have been proposed to reduce contamination of PRNU by the image content.
Kang \textit{et al.}~\cite{kang2011enhancing} proposed a method that estimates PRNU only from the phase components of noise residuals.
Lin and Li~\cite{lin2016enhancing} proposed smoothing the Fourier spectrum of the estimated PRNU using local averaging.
These two methods were based on the hypothesis that the sensor pattern noise is white noise (i.e., it has a flat frequency spectrum).
On the other hand, Li~\cite{li2010source} assume that PRNU is a weak signal, while scene details are likely to be much stronger. They proposed several models that produce weighted versions of wavelet coefficients of a noise residual.
An inverse wavelet transform on the weighted coefficients provides the enhanced noise residuals.

Various denoising filters have also been evaluated in the literature. 
These include a context-adaptive interpolator (CAI)~\cite{kang2014context}, 2-pixel approach~\cite{al2015novel}, adaptive spatial (AS) filtering~\cite{cooper2013improved}, content adaptive guided image (CAGI) filtering~\cite{zeng2016fast}, and block-matching and 3D (BM3D) algorithm~\cite{dabov2007image,chierchia2010influence}.
A recent method proposed usage of a convolutional neural network (CNN) to improve the estimation of noise residual extracted by traditional means~\cite{kirchner2020spn}.

The large size and random nature of sensor fingerprints combined with a computationally expensive fingerprint matching create difficulty in managing large databases.
Many methods have proposed alternative representations~\cite{bayram2014sensor,valsesia2015compressed,bondi2018improving,taspinar2017fast} of the PRNU fingerprint.
A recent method attempts to use a fused CNN-based approach that combines camera model features and PRNU to improve small-scale tampering detection~\cite{poyraz2020fusion}.

Despite addressing a sensor-level phenomenon, the analysis is typically performed in the RGB color space. As a result, the fingerprint is affected by various steps in the camera's ISP, e.g., demosaicing, denoising, or tone-mapping. Some researchers have proposed exploiting knowledge of specific camera components to estimate a better fingerprint, e.g., given the color filter array, one can consider only the measured color channels~\cite{li2011color}. Alternatively, a more recent line of work explores inversion of the ISP and estimation at the raw  level~\cite{mehrish2016sensor,mehrish2018robust}.

\section{Experimental Results and Discussion}
\label{sec:effect_pipes}

To assess the impact of the image processing pipeline (ISP) on PRNU fingerprints, we construct a new dataset derived from \emph{raw} images from 4 cameras and 13 different ISPs. We then follow the standard fingerprint estimation procedure~\cite{chen2008determining} and assess the impact on: (1) the correlation between different fingerprint estimates; (2) the distributions of PCE scores obtained with mismatched estimation and test ISPs; (3) the receiver operation statistics and relevant detection metrics (AUC and TPR at FPR=0.5\%).

\begin{table*}[htb!]
\centering
\caption{Dimensions of processed images obtained using all pipelines in our dataset}
\label{tab:ISP_Dataset}
\resizebox{\textwidth}{!}{
        \begin{tabular}{cccccccccccccc}
    \toprule
    {\color[HTML]{9A0000} }                                  & \multicolumn{13}{c}{{\color[HTML]{3531FF} \textbf{Camera Processing Pipelines}}}                                                                                                                                                                                                                                                  \\ \cline{2-14} \\
    \multirow{-2}{*}{{\color[HTML]{9A0000} \textbf{Camera}}} & {\color[HTML]{3531FF} \textbf{RT}} & {\color[HTML]{3531FF} \textbf{AT}} & {\color[HTML]{3531FF} \textbf{LZ}} & {\color[HTML]{3531FF} \textbf{DR}} & {\color[HTML]{3531FF} \textbf{LR}} & {\color[HTML]{3531FF} \textbf{AF}} & {\color[HTML]{3531FF} \textbf{LM}} & {\color[HTML]{3531FF} \textbf{DX}} & {\color[HTML]{3531FF} \textbf{C1}} & {\color[HTML]{3531FF} \textbf{LT}} & {\color[HTML]{3531FF} \textbf{IN}} & {\color[HTML]{3531FF} \textbf{UN}} & {\color[HTML]{3531FF} \textbf{DN}} \\ \midrule   
    {\color[HTML]{9A0000} \textbf{N7k}}                       & 3272 x 4940                        & 3264 x 4928                        & 3270 x 4938                        & 3280 x 4948                        & 3280 x 4948                        & 3280 x 4948 & 3264 x 4928 & 3264 x 4928                        & Diff. sizes                        & 3264 x 4928                        & 3280 x 4948                        & 3280 x 4948                        & 3280 x 4948                                                  \\ 
    {\color[HTML]{9A0000} \textbf{N90}}                       & 2860 x 4302                        & 2860 x 4290                        & 2858 x 4300                        & 2868 x 4310                        & 2868 x 4310                        & 2868 x 4310 & 2848 x 4288 & 2848 x 4288                        & Diff. sizes                        & 2848 x 4288                        & 2868 x 4310                        & 2868 x 4310                        & 2868 x 4310                               \\ 
    {\color[HTML]{9A0000} \textbf{C4D}}                      & 2594 x 3900                        & -                                  & 2592 x 3898                        & 2602 x 3908                        & 2602 x 3908                        & 2602 x 3908 & 2592 x 3888 & 2592 x 3888                        & 2592 x 3888                        & 2592 x 3888                        & 2602 x 3908                        & 2602 x 3908                        & 2602 x 3908                               \\ 
    {\color[HTML]{9A0000} \textbf{C5D}}                      & 2912 x 4378                        & -                                  & 2910 x 4376                        & 2920 x 4386                        & 2920 x 4386                        & 2920 x 4386 & 2912 x 4368 & 2912 x 4368                        & 2912 x 4368                        & Diff. sizes                        & 2920 x 4386                        & 2920 x 4386                        & 2920 x 4386
                               \\ \bottomrule
    \end{tabular}
}
\end{table*}
\begin{table}[t!]
    \centering
    \caption{Correlation coefficients between PRNU fingerprints using different imaging pipelines.}
    \label{tab:ISP_Corr_N7}
    \resizebox{0.49\textwidth}{!}{
\begin{tabular}{cccccccccccccc}
    \toprule
                                    
     & {\color[HTML]{3531FF} \textbf{RT}} & {\color[HTML]{3531FF} \textbf{AT}} & {\color[HTML]{3531FF} \textbf{LZ}} & {\color[HTML]{3531FF} \textbf{DR}} & {\color[HTML]{3531FF} \textbf{LR}} & {\color[HTML]{3531FF} \textbf{AF}} & {\color[HTML]{3531FF} \textbf{LM}} & {\color[HTML]{3531FF} \textbf{DX}} & {\color[HTML]{3531FF} \textbf{C1}} & {\color[HTML]{3531FF} \textbf{LT}} & {\color[HTML]{3531FF} \textbf{IN}} & {\color[HTML]{3531FF} \textbf{UN}} & {\color[HTML]{3531FF} \textbf{DN}} \\ \midrule
    {\color[HTML]{9A0000} }                                                                                & \multicolumn{13}{c}{{\color[HTML]{3531FF} \textbf{Nikon D7000}}}                                                                                                                                                                                                                                                                       \\ \midrule
    
    {\color[HTML]{9A0000} \textbf{RT}}                                                                     & \textbf{1.00}                       & {\color[HTML]{FE0000} \textbf{0.03}} & {\color[HTML]{FE0000} \textbf{0.02}} & {\color[HTML]{FE0000} \textbf{0.02}} & {\color[HTML]{FE0000} \textbf{0.02}} & {\color[HTML]{FE0000} \textbf{0.02}} & {\color[HTML]{FE0000} \textbf{0.02}} & {\color[HTML]{FE0000} \textbf{0.03}} & {\color[HTML]{FE0000} \textbf{0.00}} & {\color[HTML]{FE0000} \textbf{0.03}} & {\color[HTML]{FE0000} \textbf{0.02}} & {\color[HTML]{FE0000} \textbf{0.02}} & {\color[HTML]{FE0000} \textbf{0.02}} \\ 
    {\color[HTML]{9A0000} \textbf{AT}}                                                                     & {\color[HTML]{FE0000} \textbf{0.03}} & \textbf{1.00}                       & 0.45 & 0.40 & 0.42 & 0.43 & 0.35 & 0.78                                & {\color[HTML]{FE0000} \textbf{0.01}} & 0.69 & 0.34 & 0.33 & 0.56 \\
    {\color[HTML]{9A0000} \textbf{LZ}}                                                                     & {\color[HTML]{FE0000} \textbf{0.02}} & 0.45                               & \textbf{1.00}                      & 0.65 & 0.71 & 0.71 & 0.45 & 0.50 & {\color[HTML]{FE0000} \textbf{0.00}} & 0.50 & 0.24 & 0.58 & 0.42                               \\ 
    {\color[HTML]{9A0000} \textbf{DR}}                                                                     & {\color[HTML]{FE0000} \textbf{0.02}} & 0.40 & 0.65                    & \textbf{1.00}                      & 0.89 & 0.88 & 0.44 & 0.46 & {\color[HTML]{FE0000} \textbf{0.00}} & 0.46 & 0.26 & 0.65 & 0.45                              \\ 
    {\color[HTML]{9A0000} \textbf{LR}}                                                                     & {\color[HTML]{FE0000} \textbf{0.02}} & 0.42                                 & 0.71                                 & 0.89      & \textbf{1.00}                      & 0.93                                 & 0.50                                 & 0.48                                & {\color[HTML]{FE0000} \textbf{0.00}} & 0.50                                 & 0.27                                 & 0.69                                 & 0.47                                                               \\ 
    {\color[HTML]{9A0000} \textbf{AF}}                                                                     & {\color[HTML]{FE0000} \textbf{0.02}} & 0.43                                 & 0.71                                 & 0.88                                 & 0.93  & \textbf{1.00}                      & 0.50                                 & 0.49                      & {\color[HTML]{FE0000} \textbf{0.00}} & 0.51                                 & 0.28                                 & 0.67                                 & 0.48     \\ 
    {\color[HTML]{9A0000} \textbf{LM}}                                                                     & {\color[HTML]{FE0000} \textbf{0.02}} & 0.35                                 & 0.45                                 & 0.44                                 & 0.50                                 & 0.50                           & \textbf{1.00}                      & 0.44                               & {\color[HTML]{FE0000} \textbf{0.00}} & 0.51                                 & 0.28                                 & 0.44                                 & 0.44                                 \\
    {\color[HTML]{9A0000} \textbf{DX}} & {\color[HTML]{FE0000} \textbf{0.03}} & 0.78                                 & 0.50                                 & 0.46                                 & 0.48                                 & 0.49                                 & 0.44                                 & \textbf{1.00}                        & {\color[HTML]{FE0000} \textbf{0.01}} & 0.84                                 & 0.37                                 & 0.39                                 & 0.64 \\ 
     {\color[HTML]{9A0000} \textbf{C1}} & {\color[HTML]{FE0000} \textbf{0.00}} & {\color[HTML]{FE0000} \textbf{0.01}} & {\color[HTML]{FE0000} \textbf{0.00}} & {\color[HTML]{FE0000} \textbf{0.00}} & {\color[HTML]{FE0000} \textbf{0.00}} & {\color[HTML]{FE0000} \textbf{0.00}} & {\color[HTML]{FE0000} \textbf{0.00}} & {\color[HTML]{FE0000} \textbf{0.01}} & \textbf{1.00}                        & {\color[HTML]{FE0000} \textbf{0.01}} & {\color[HTML]{FE0000} \textbf{0.01}} & {\color[HTML]{FE0000} \textbf{0.00}} & {\color[HTML]{FE0000} \textbf{0.01}} \\ 
    {\color[HTML]{9A0000} \textbf{LT}}                                                                     & {\color[HTML]{FE0000} \textbf{0.03}} & 0.69                                 & 0.50                                 & 0.46                                 & 0.50                                 & 0.51                                 & 0.51                                 & 0.84                                 & {\color[HTML]{FE0000} \textbf{0.01}} & \textbf{1.00}                        & 0.37                                 & 0.40                                 & 0.65                                 \\ 
    {\color[HTML]{9A0000} \textbf{IN}} & {\color[HTML]{FE0000} \textbf{0.02}} & 0.34                                 & 0.24                                 & 0.26                                 & 0.27                                 & 0.28                                 & 0.28                                 & 0.37                                 & {\color[HTML]{FE0000} \textbf{0.01}} & 0.37                                 & \textbf{1.00}                        & 0.28                                 & 0.47                                 \\ 
    {\color[HTML]{9A0000} \textbf{UN}} & {\color[HTML]{FE0000} \textbf{0.02}} & 0.33                                 & 0.58                                 & 0.65                                 & 0.69                                 & 0.67                                 & 0.44                                 & 0.39                                 & {\color[HTML]{FE0000} \textbf{0.00}} & 0.40                                 & 0.28                                 & \textbf{1.00}                        & 0.41                                 \\ 
    {\color[HTML]{9A0000} \textbf{DN}} & {\color[HTML]{FE0000} \textbf{0.02}} & 0.56                                 & 0.42                                 & 0.45                                 & 0.47                                 & 0.48                                 & 0.44                                 & 0.64                                 & {\color[HTML]{FE0000} \textbf{0.01}} & 0.65                                 & 0.47                                 & 0.41                                 & \textbf{1.00}                        \\ 
    \multicolumn{1}{l}{}                                                                                 & \multicolumn{13}{c}{{\color[HTML]{3531FF} \textbf{Nikon D90}}}                                                                                                                                                                                                                                                                                                                                                                \\ \midrule
    {\color[HTML]{9A0000} \textbf{RT}}  & \textbf{1.00}                        & {\color[HTML]{FE0000} \textbf{0.01}} & {\color[HTML]{FE0000} \textbf{0.01}} & {\color[HTML]{FE0000} \textbf{0.01}} & {\color[HTML]{FE0000} \textbf{0.01}} & {\color[HTML]{FE0000} \textbf{0.01}} & {\color[HTML]{FE0000} \textbf{0.01}} & {\color[HTML]{FE0000} \textbf{0.01}} & {\color[HTML]{FE0000} \textbf{0.00}} & {\color[HTML]{FE0000} \textbf{0.01}} & {\color[HTML]{FE0000} \textbf{0.01}} & {\color[HTML]{FE0000} \textbf{0.01}} & {\color[HTML]{FE0000} \textbf{0.01}} \\
{\color[HTML]{9A0000} \textbf{AT}}  & {\color[HTML]{FE0000} \textbf{0.01}} & \textbf{1.00}                        & 0.32                                 & 0.28                                 & 0.30                                 & 0.29                                 & 0.28                                 & 0.39                                 & {\color[HTML]{FE0000} \textbf{0.00}} & 0.35                                 & 0.15                                 & 0.26                                 & 0.23                                 \\
{\color[HTML]{9A0000} \textbf{LZ}}  & {\color[HTML]{FE0000} \textbf{0.01}} & 0.32                                 & \textbf{1.00}                        & 0.75                                 & 0.77                                 & 0.79                                 & 0.39                                 & 0.41                                 & {\color[HTML]{FE0000} \textbf{0.01}} & 0.43                                 & 0.15                                 & 0.62                                 & 0.29                                 \\
{\color[HTML]{9A0000} \textbf{DR}}  & {\color[HTML]{FE0000} \textbf{0.01}} & 0.28                                 & 0.75                                 & \textbf{1.00}                        & 0.83                                 & 0.88                                 & 0.33                                 & 0.35                                 & {\color[HTML]{FE0000} \textbf{0.00}} & 0.36                                 & 0.14                                 & 0.58                                 & 0.28                                 \\
{\color[HTML]{9A0000} \textbf{LR}}  & {\color[HTML]{FE0000} \textbf{0.01}} & 0.30                                 & 0.77                                 & 0.83                                 & \textbf{1.00}                        & 0.89                                 & 0.39                                 & 0.40                                 & {\color[HTML]{FE0000} \textbf{0.01}} & 0.42                                 & 0.14                                 & 0.64                                 & 0.31                                 \\
{\color[HTML]{9A0000} \textbf{AF}}  & {\color[HTML]{FE0000} \textbf{0.01}} & 0.29                                 & 0.79                                 & 0.88                                 & 0.89                                 & \textbf{1.00}                        & 0.37                                 & 0.37                                 & {\color[HTML]{FE0000} \textbf{0.01}} & 0.4                                  & 0.13                                 & 0.6                                  & 0.29                                 \\
{\color[HTML]{9A0000} \textbf{LM}}  & {\color[HTML]{FE0000} \textbf{0.01}} & 0.28                                 & 0.39                                 & 0.33                                 & 0.39                                 & 0.37                                 & \textbf{1.00}                        & 0.7                                  & {\color[HTML]{FE0000} \textbf{0.01}} & 0.44                                 & 0.11                                 & 0.31                                 & 0.27                                 \\
{\color[HTML]{9A0000} \textbf{DX}}  & {\color[HTML]{FE0000} \textbf{0.01}} & 0.39                                 & 0.41                                 & 0.35                                 & 0.40                                 & 0.37                                 & 0.70                                 & \textbf{1.00}                        & {\color[HTML]{FE0000} \textbf{0.01}} & 0.47                                 & 0.17                                 & 0.33                                 & 0.30                                 \\
{\color[HTML]{9A0000} \textbf{C1}}  & {\color[HTML]{FE0000} \textbf{0.00}} & {\color[HTML]{FE0000} \textbf{0.00}} & {\color[HTML]{FE0000} \textbf{0.01}} & {\color[HTML]{FE0000} \textbf{0.00}} & {\color[HTML]{FE0000} \textbf{0.01}} & {\color[HTML]{FE0000} \textbf{0.01}} & {\color[HTML]{FE0000} \textbf{0.01}} & {\color[HTML]{FE0000} \textbf{0.01}} & \textbf{1.00}                        & {\color[HTML]{FE0000} \textbf{0.01}} & {\color[HTML]{FE0000} \textbf{0.01}} & {\color[HTML]{FE0000} \textbf{0.00}} & {\color[HTML]{FE0000} \textbf{0.01}} \\
{\color[HTML]{9A0000} \textbf{LT}}  & {\color[HTML]{FE0000} \textbf{0.01}} & 0.35                                 & 0.43                                 & 0.36                                 & 0.42                                 & 0.40                                 & 0.44                                 & 0.47                                 & {\color[HTML]{FE0000} \textbf{0.01}} & \textbf{1.00}                        & 0.25                                 & 0.33                                 & 0.48                                 \\
{\color[HTML]{9A0000} \textbf{IN}}  & {\color[HTML]{FE0000} \textbf{0.01}} & 0.15                                 & 0.15                                 & 0.14                                 & 0.14                                 & 0.13                                 & 0.11                                 & 0.17                                 & {\color[HTML]{FE0000} \textbf{0.01}} & 0.25                                 & \textbf{1.00}                        & 0.17                                 & 0.38                                 \\
{\color[HTML]{9A0000} \textbf{UN}}  & {\color[HTML]{FE0000} \textbf{0.01}} & 0.26                                 & 0.62                                 & 0.58                                 & 0.64                                 & 0.60                                 & 0.31                                 & 0.33                                 & {\color[HTML]{FE0000} \textbf{0.00}} & 0.33                                 & 0.17                                 & \textbf{1.00}                        & 0.31                                 \\
{\color[HTML]{9A0000} \textbf{DN}}  & {\color[HTML]{FE0000} \textbf{0.01}} & 0.23                                 & 0.29                                 & 0.28                                 & 0.31                                 & 0.29                                 & 0.27                                 & 0.30                                 & {\color[HTML]{FE0000} \textbf{0.01}} & 0.48                                 & 0.38                                 & 0.31                                 & \textbf{1.00}                        \\
\multicolumn{1}{l}{} &
\multicolumn{13}{c}{{\color[HTML]{3531FF} \textbf{Canon EOS 40D}}} \\ \midrule      
    {\color[HTML]{9A0000} \textbf{RT}}  & \textbf{1.00}                        & -                                    & 0.69                                 & 0.74                                 & 0.74                                 & 0.75                                 & 0.70                                 & 0.72                                 & 0.51                                 & 0.71                                 & 0.24                                 & 0.16                                 & 0.59                                 \\
{\color[HTML]{9A0000} \textbf{AT}}  & -                                    & -                                    & -                                    & -                                    & -                                    & -                                    & -                                    & -                                    & -                                    & -                                    & -                                    & -                                    & -                                    \\
{\color[HTML]{9A0000} \textbf{LZ}}  & 0.69                                 & -                                    & \textbf{1.00}                        & 0.65                                 & 0.72                                 & 0.71                                 & 0.68                                 & 0.82                                 & 0.54                                 & 0.82                                 & 0.24                                 & 0.14                                 & 0.57                                 \\
{\color[HTML]{9A0000} \textbf{DR}}  & 0.74                                 & -                                    & 0.65                                 & \textbf{1.00}                        & 0.84                                 & 0.84                                 & 0.63                                 & 0.67                                 & 0.45                                 & 0.64                                 & 0.23                                 & 0.17                                 & 0.60                                 \\
{\color[HTML]{9A0000} \textbf{LR}}  & 0.74                                 & -                                    & 0.72                                 & 0.84                                 & \textbf{1.00}                        & 0.89                                 & 0.71                                 & 0.74                                 & 0.53                                 & 0.72                                 & 0.25                                 & 0.16                                 & 0.66                                 \\
{\color[HTML]{9A0000} \textbf{AF}}  & 0.75                                 & -                                    & 0.71                                 & 0.84                                 & 0.89                                 & \textbf{1.00}                        & 0.75                                 & 0.71                                 & 0.52                                 & 0.69                                 & 0.24                                 & 0.15                                 & 0.63                                 \\
{\color[HTML]{9A0000} \textbf{LM}}  & 0.70                                 & -                                    & 0.68                                 & 0.63                                 & 0.71                                 & 0.75                                 & \textbf{1.00}                        & 0.67                                 & 0.57                                 & 0.71                                 & 0.21                                 & 0.12                                 & 0.58                                 \\
{\color[HTML]{9A0000} \textbf{DX}}  & 0.72                                 & -                                    & 0.82                                 & 0.67                                 & 0.74                                 & 0.71                                 & 0.67                                 & \textbf{1.00}                        & 0.55                                 & 0.84                                 & 0.28                                 & 0.15                                 & 0.60                                 \\
{\color[HTML]{9A0000} \textbf{C1}}  & 0.51                                 & -                                    & 0.54                                 & 0.45                                 & 0.53                                 & 0.52                                 & 0.57                                 & 0.55                                 & \textbf{1.00}                        & 0.55                                 & 0.17                                 & 0.10                                 & 0.42                                 \\
{\color[HTML]{9A0000} \textbf{LT}}  & 0.71                                 & -                                    & 0.82                                 & 0.64                                 & 0.72                                 & 0.69                                 & 0.71                                 & 0.84                                 & 0.55                                 & \textbf{1.00}                        & 0.25                                 & 0.15                                 & 0.57                                 \\
{\color[HTML]{9A0000} \textbf{IN}}  & 0.24                                 & -                                    & 0.24                                 & 0.23                                 & 0.25                                 & 0.24                                 & 0.21                                 & 0.28                                 & 0.17                                 & 0.25                                 & \textbf{1.00}                        & 0.21                                 & 0.26                                 \\
{\color[HTML]{9A0000} \textbf{UN}}  & 0.16                                 & -                                    & 0.14                                 & 0.17                                 & 0.16                                 & 0.15                                 & 0.12                                 & 0.15                                 & 0.10                                 & 0.15                                 & 0.21                                 & \textbf{1.00}                        & 0.17                                 \\
{\color[HTML]{9A0000} \textbf{DN}}  & 0.59                                 & -                                    & 0.57                                 & 0.60                                 & 0.66                                 & 0.63                                 & 0.58                                 & 0.60                                 & 0.42                                 & 0.57                                 & 0.26                                 & 0.17                                 & \textbf{1.00}                        \\
    \multicolumn{1}{l}{}     &                    \multicolumn{13}{c}{{\color[HTML]{3531FF} \textbf{Canon EOS 5D}}}                             \\ \midrule
    {\color[HTML]{9A0000} \textbf{RT}}  & \textbf{1.00}                        & -                                    & 0.39                                 & 0.39                                 & 0.40                                 & 0.40                                 & 0.61                                 & 0.32                                 & 0.46                                 & 0.60                                 & 0.28                                 & 0.29                                 & 0.59                                 \\
{\color[HTML]{9A0000} \textbf{AT}}  & -                                    & -                                    & -                                    & -                                    & -                                    & -                                    & -                                    & -                                    & -                                    & -                                    & -                                    & -                                    & -                                    \\
{\color[HTML]{9A0000} \textbf{LZ}}  & 0.39                                 & -                                    & \textbf{1.00}                        & 0.68                                 & 0.73                                 & 0.74                                 & 0.30                                 & 0.64                                 & 0.24                                 & 0.33                                 & 0.15                                 & 0.55                                 & 0.32                                 \\
{\color[HTML]{9A0000} \textbf{DR}}  & 0.39                                 & -                                    & 0.68                                 & \textbf{1.00}                        & 0.81                                 & 0.87                                 & 0.28                                 & 0.55                                 & 0.21                                 & 0.28                                 & 0.14                                 & 0.54                                 & 0.32                                 \\
{\color[HTML]{9A0000} \textbf{LR}}  & 0.40                                 & -                                    & 0.73                                 & 0.81                                 & \textbf{1.00}                        & 0.88                                 & 0.31                                 & 0.61                                 & 0.24                                 & 0.33                                 & 0.15                                 & 0.59                                 & 0.35                                 \\
{\color[HTML]{9A0000} \textbf{AF}}  & 0.40                                 & -                                    & 0.74                                 & 0.87                                 & 0.88                                 & \textbf{1.00}                        & 0.31                                 & 0.59                                 & 0.24                                 & 0.31                                 & 0.14                                 & 0.57                                 & 0.34                                 \\
{\color[HTML]{9A0000} \textbf{LM}}  & 0.61                                 & -                                    & 0.30                                 & 0.28                                 & 0.31                                 & 0.31                                 & \textbf{1.00}                        & 0.36                                 & 0.58                                 & 0.72                                 & 0.15                                 & 0.24                                 & 0.51                                 \\
{\color[HTML]{9A0000} \textbf{DX}}  & 0.32                                 & -                                    & 0.64                                 & 0.55                                 & 0.61                                 & 0.59                                 & 0.36                                 & \textbf{1.00}                        & 0.30                                 & 0.43                                 & 0.12                                 & 0.40                                 & 0.27                                 \\
{\color[HTML]{9A0000} \textbf{C1}}  & 0.46                                 & -                                    & 0.24                                 & 0.21                                 & 0.24                                 & 0.24                                 & 0.58                                 & 0.30                                 & \textbf{1.00}                        & 0.56                                 & 0.13                                 & 0.17                                 & 0.34                                 \\
{\color[HTML]{9A0000} \textbf{LT}}  & 0.60                                 & -                                    & 0.33                                 & 0.28                                 & 0.33                                 & 0.31                                 & 0.72                                 & 0.43                                 & 0.56                                 & \textbf{1.00}                        & 0.20                                 & 0.22                                 & 0.46                                 \\
{\color[HTML]{9A0000} \textbf{IN}}  & 0.28                                 & -                                    & 0.15                                 & 0.14                                 & 0.15                                 & 0.14                                 & 0.15                                 & 0.12                                 & 0.13                                 & 0.20                                 & \textbf{1.00}                        & 0.12                                 & 0.25                                 \\
{\color[HTML]{9A0000} \textbf{UN}}  & 0.29                                 & -                                    & 0.55                                 & 0.54                                 & 0.59                                 & 0.57                                 & 0.24                                 & 0.4                                  & 0.17                                 & 0.22                                 & 0.12                                 & \textbf{1.00}                        & 0.30                                 \\
{\color[HTML]{9A0000} \textbf{DN}}  & 0.59                                 & -                                    & 0.32                                 & 0.32                                 & 0.35                                 & 0.34                                 & 0.51                                 & 0.27                                 & 0.34                                 & 0.46                                 & 0.25                                 & 0.30                                 & \textbf{1.00}                        \\  \bottomrule
    \end{tabular}
    }
    \end{table}

\subsection{Dataset and Fingerprint Estimation}
\label{subsec:ISP_Dataset}
We collected \emph{raw} images for 4 cameras: Nikon D7000 (N7k), Nikon D90 (N90), Canon EOS 40D (C4D) and Canon EOS 5D (C5D). The Nikon and Canon images were taken from the Raise~\cite{dang2015raise}, and MIT-5k~\cite{bychkovsky2011learning} datasets, respectively (Table~\ref{tab:ISP_Dataset}). We collected  120 full-resolution images, out of which 60 were used for PRNU estimation and 60 for subsequent attribution experiments. 
% {\color{green}Are you sure? I have 150 images per camera in my toolbox - why would we ignore 30 images?} {\color{purple} - We kept aside 30 images as you had something in mind at that time, I don't remember the exact thing}

We then processed all images using 13 different ISPs. We included popular digital darkroom software (both commercial and open-source) as well as recent neural-network-based ISPs. We included 10 darkrooms:  RawTherapee 5.6~(RT)~\cite{bib:rawtherapee}, Corel AfterShot Pro 3~(AT)~\cite{bib:aftershotpro} LightZone 4.1.9~(LZ)~\cite{bib:LightZone}, DCRaw 9.27~(DR)~\cite{bib:dcraw}, LibRaw 0.17.2~(LR)~\cite{bib:libraw}, Affinity Photo 1.6.6~(AF)~\cite{bib:affinity}, Luminar 3~(LR)~\cite{bib:skylum}, DXO PhotoLab 2~(DX)~\cite{bib:dxomark}, Capture One 12~(C1)~\cite{bib:captureone}, and Adobe Lightroom 5~(LT)~\cite{bib:lightroom}. The neural ISPs included a popular model for joint demosaicing and denoising (DNet)~\cite{gharbi2016deep}, the UNet~(UN)~\cite{ronneberger2015u} model, and a simple network that mimics a standard ISP (INet)~\cite{korus2019neural}. All NN models were trained to reproduce the result of a standard imaging pipeline.

To assess the most optimistic scenario, we made sure to: (1) select the same images for PRNU estimation; (2) use an \emph{automatic} mode with default settings and no scaling in each ISP; (3) work on uncompressed bitmap images. In a real-world setting, these conditions are unlikely to be met.

\subsection{Analysis of Camera Fingerprint Similarity}
\label{subsec:ISP_PRNUCorr}

To compare the PRNU fingerprints, we compute the normalized correlation coefficients between all possible pairs of estimation pipelines. If necessary, we align the fingerprints by shifting one of them to the location indicated by a peak in cross-correlation (needed sometimes due to minor differences in how ISPs crop the \emph{raw} images; see Table V in the supplemental file).
% \footnote{Supplemental file is available on author's website and arXiv:2004.01929.}).

We collected the obtained results in Table~\ref{tab:ISP_Corr_N7} for all four cameras.
% for Nikon D7000; the results are similar for other cameras). 
It can be observed that despite a conservative evaluation setting, the estimated fingerprints tend to differ considerably. The average correlation for all cross-ISP pairs is only 0.38. Some of the ISPs, e.g., DCRaw, LibRaw, and Affinity, tend to correlate better, which may indicate some common components in their functionality (e.g., use of the DCRaw routines in LibRaw~\cite{bib:dcraw_libraw}).

Interestingly, fingerprints from \emph{CaptureOne} and \emph{RawTherapee} do not correlate with any other ISPs for neither of Nikon cameras (marked in red). Based on visual comparison of image patches, it would appear both programs apply some slight geometric transformation while cropping full-frame \emph{raw} images, which de-synchronizes the fingerprints. We emphasize that this happens despite explicit instructions to leave image size intact and that the problem does not occur for Canon cameras (this may be related to the format of \emph{raw} images in our dataset - NEF for Nikon and DNG for Canon cameras). 

% (results are presented in Section I of supplementary material)
\begin{figure}[t!]
    \centering
    \includegraphics[width=\columnwidth] {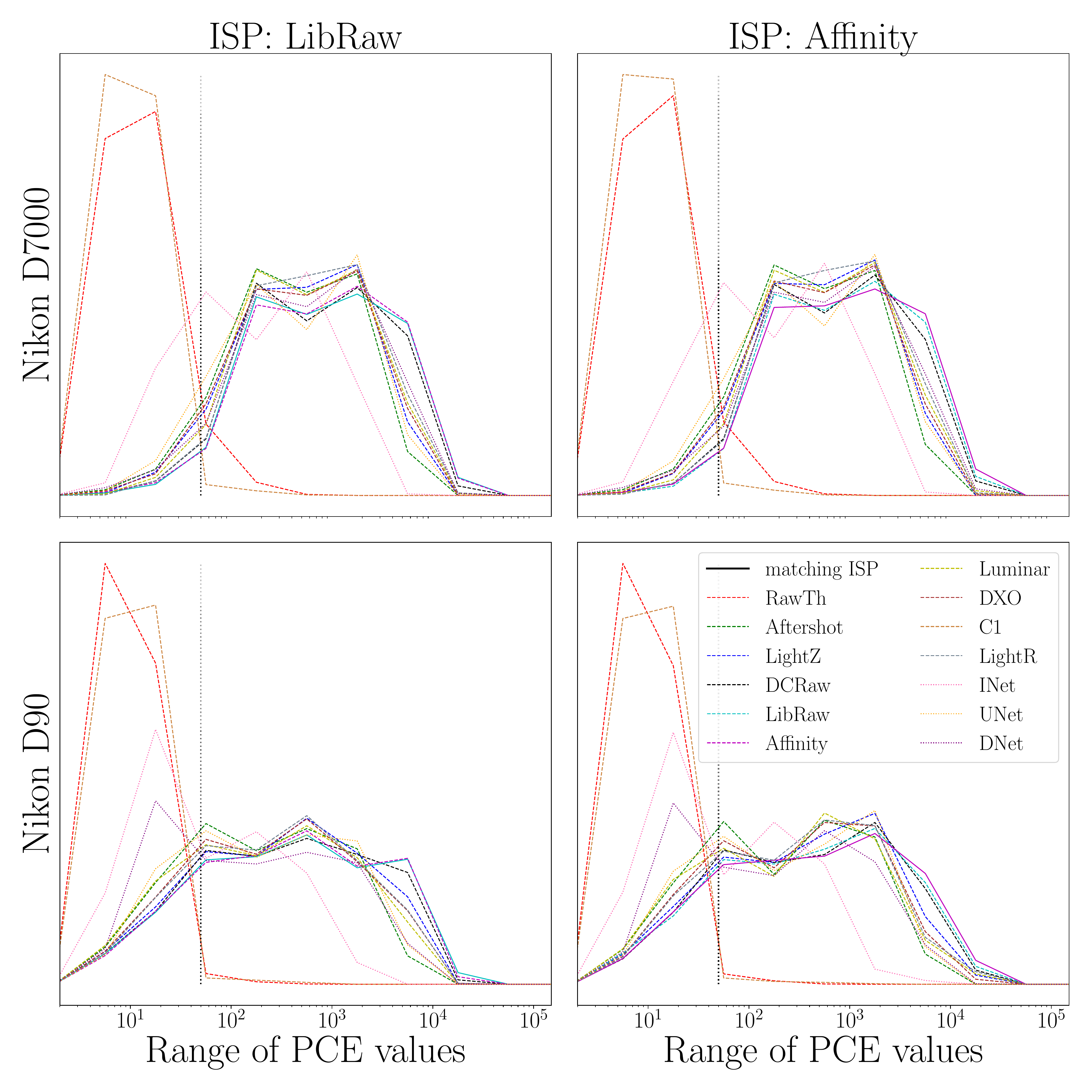}
    \caption{Distribution of PCE scores for cross-ISP matching (512~px patches); the dashed vertical line shows a common acceptance threshold of 50.}
    % {\color{red} why are some lines nearly flat?} 
    % {\color{blue}- No Idea at this point} {\color{red}I checked counts in the code, and they seem to have less samples - why?} {\color{blue}- As mentioned in my last email, these results were not using multiple patches, I will email the updated pces which has multiple patches for all pipes} {\color{red} also, please double check the failed matches for neural ISPs for Canon cameras - do we see the same thing in PCE distribution tables (in supplement?)?} {\color{blue}- emailed the supplement}}
    \label{fig:ISP_PCEDist_Libraw}
\end{figure}

\begin{figure}[t!]
    \centering
    \includegraphics[width=\columnwidth]{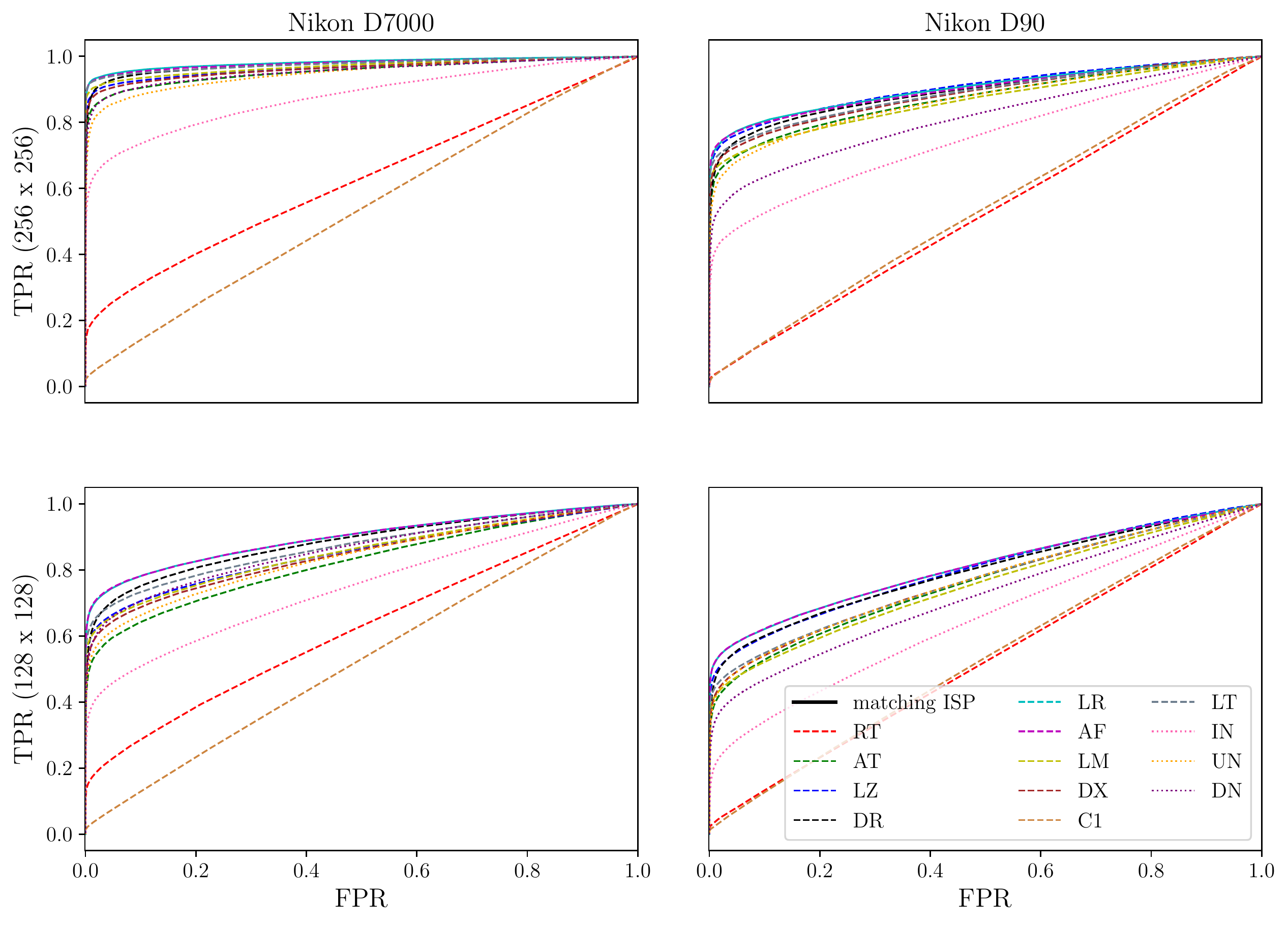}
    \caption{Receiver operating characteristic (ROC) curves with LibRaw as PRNU estimation ISP and all pipelines as test.}
    \label{fig:ISP_ROC_PatchSizes}
\end{figure}

\begin{figure*}[t!]
    \centering
    \includegraphics[width=\textwidth,trim={0 0 24in 0},clip] {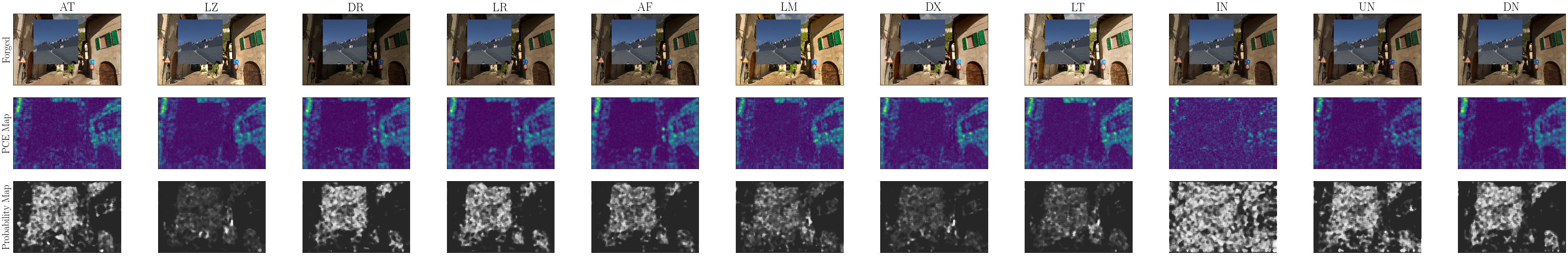} \\ %l,b,r,t
    \vspace{0.1in}
    \includegraphics[width=0.85\textwidth,trim={29in 0 0 0},clip] {Images/Tamper_1_horz_overlap32_norm_color.png}
    \caption{Example tampering localization results obtained with a camera fingerprint from a mismatched ISP (LibRAW); tampering probability maps were obtained as \emph{p}-values from PCE statistics and were post-processed for presentation clarity (contrast + median filter).}
    \label{fig:ISP_localization}
\end{figure*}

\subsection{Impact on PCE Detection Statistics}
\label{subsec:ISP_PCEDist}

To assess the impact on fingerprint detection statistics, we measured peak-to-correlation-energy (PCE) for all possible pairs of ISPs (using different pipelines for PRNU estimation and test images). To increase the number of samples, we extract all possible non-overlapping patches from full-resolution images. This yields from 600 to 55,000 patches for patch sizes ranging from 1024~px to 128~px. 

In general, using the same imaging pipeline leads to the best matching performance (see median PCE scores for Nikon D7000 in Table~\ref{tab:ISP_PCE_Med_N7} and Canon EOS 40 D in Table~\ref{tab:ISP_PCE_Med_C4} of the supplemental file). In cross-ISP matching, the PCE deteriorates on average by 61, 56, 50, and 41\% for patches of 1024, 512, 256, and 128 px, respectively. Full distributions of PCE scores for 512~px patches from selected cameras and ISPs are shown in Figure~\ref{fig:ISP_PCEDist_Libraw} (results for all cameras are depicted in~Figure~\ref{fig:ISP_PCEDist_Libraw_Supp} of the supplemental file). Note that the PCE distributions are shown in logarithmic scale. Configurations with the same ISP are shown with solid lines, whereas mismatched digital darkrooms and neural ISPs are shown with dashed and dotted lines, respectively. 

Overall, we see a similar trend as in fingerprint correlation (Section~\ref{subsec:ISP_PRNUCorr}). Apart from obvious matching failures for \emph{CaptureOne} and \emph{RawTherapee} for Nikon cameras, we observe various degrees of PCE deterioration. The most significant and consistent deterioration occurs for neural ISPs. In some cases, e.g., for Canon cameras in our experiments, they seemed to have incompatible fingerprints.

\begin{table}[t!]
    \centering
    \caption{Median PCE values for cross-pipeline attribution of non-overlapping patches from Nikon D7000}
    \label{tab:ISP_PCE_Med_N7}
    \resizebox{0.49\textwidth}{!}{
    \begin{tabular}{cccccccccccccc}
    \toprule
    {\color[HTML]{9A0000} }                                                                                & \multicolumn{13}{c}{{\color[HTML]{3531FF} \textbf{Image Processing Pipelines for Test Images}}}                                                                                                                                                                                                                                                                       \\ \cline{2-14} \\
    \multirow{-2}{*}{{\color[HTML]{9A0000} \textbf{\begin{tabular}[c]{@{}c@{}}PRNU \\ ISP\end{tabular}}}} & {\color[HTML]{3531FF} \textbf{RT}} & {\color[HTML]{3531FF} \textbf{AT}} & {\color[HTML]{3531FF} \textbf{LZ}} & {\color[HTML]{3531FF} \textbf{DR}} & {\color[HTML]{3531FF} \textbf{LR}} & {\color[HTML]{3531FF} \textbf{AF}} & {\color[HTML]{3531FF} \textbf{LM}} & {\color[HTML]{3531FF} \textbf{DX}} & {\color[HTML]{3531FF} \textbf{C1}} & {\color[HTML]{3531FF} \textbf{LT}} & {\color[HTML]{3531FF} \textbf{IN}} & {\color[HTML]{3531FF} \textbf{UN}} & {\color[HTML]{3531FF} \textbf{DN}} \\ \midrule
    {\color[HTML]{9A0000} }                                                                                & \multicolumn{13}{c}{{\color[HTML]{3531FF} \textbf{Patch size = 1024 x 1024 pixels ($\sim$600 patches per camera per pipeline)}}}                                                                                                                                                                                                                                                                       \\ \midrule
    
    {\color[HTML]{9A0000} \textbf{RT}}                                                                     & \textbf{198}                       & {\color[HTML]{FE0000} \textbf{10}} & {\color[HTML]{FE0000} \textbf{10}} & {\color[HTML]{FE0000} \textbf{11}} & {\color[HTML]{FE0000} \textbf{11}} & {\color[HTML]{FE0000} \textbf{11}} & {\color[HTML]{FE0000} \textbf{11}} & {\color[HTML]{FE0000} \textbf{10}} & {\color[HTML]{FE0000} \textbf{10}} & {\color[HTML]{FE0000} \textbf{10}} & {\color[HTML]{FE0000} \textbf{11}} & {\color[HTML]{FE0000} \textbf{12}} & {\color[HTML]{FE0000} \textbf{13}} \\ 
    {\color[HTML]{9A0000} \textbf{AT}}                                                                     & {\color[HTML]{FE0000} \textbf{10}} & \textbf{880}                       & 865                                & 978                                & 1105                               & 1176                               & 833                                & 936                                & {\color[HTML]{FE0000} \textbf{10}} & 1072                               & 209                                & 550                                & 926                                \\ 
    {\color[HTML]{9A0000} \textbf{LZ}}                                                                     & {\color[HTML]{FE0000} \textbf{11}} & 1617                               & \textbf{2324}                      & 2338                               & 2714                               & 3057                               & 2333                               & 2071                               & {\color[HTML]{FE0000} \textbf{10}} & 2584                               & 632                                & 1523                               & 2326                               \\ 
    {\color[HTML]{9A0000} \textbf{DR}}                                                                     & {\color[HTML]{FE0000} \textbf{11}} & 1591                               & 2042                               & \textbf{2840}                      & 3169                               & 3389                               & 2355                               & 1982                               & {\color[HTML]{FE0000} \textbf{10}} & 2539                               & 675                                & 1838                               & 2663                               \\ 
    {\color[HTML]{9A0000} \textbf{LR}}                                                                     & {\color[HTML]{FE0000} \textbf{11}} & 1766                               & 2310                               & 3060                               & \textbf{3538}                      & 3802                               & 2647                               & 2242                               & {\color[HTML]{FE0000} \textbf{10}} & 2786                               & 769                                & 2049                               & 3036                               \\ 
    {\color[HTML]{9A0000} \textbf{AF}}                                                                     & {\color[HTML]{FE0000} \textbf{10}} & 1672                               & 2137                               & 2770                               & 3300                               & \textbf{3626}                      & 2520                               & 2113                               & {\color[HTML]{FE0000} \textbf{10}} & 2719                               & 743                                & 1843                               & 2721                               \\ 
    {\color[HTML]{9A0000} \textbf{LM}}                                                                     & {\color[HTML]{FE0000} \textbf{12}} & 1915                               & 2692                               & 3635                               & 4252                               & 4355                               & \textbf{5098}                      & 2977                               & {\color[HTML]{FE0000} \textbf{10}} & 3833                               & 1255                               & 2923                               & 3741                               \\ 
    {\color[HTML]{9A0000} \textbf{DX}}                                                                     & {\color[HTML]{FE0000} \textbf{10}} & 926                                & 1046                               & 1187                               & 1385                               & 1471                               & 1233                               & \textbf{1170}                      & {\color[HTML]{FE0000} \textbf{10}} & 1436                               & 224                                & 699                                & 1183                               \\ 
    {\color[HTML]{9A0000} \textbf{C1}}                                                                     & {\color[HTML]{FE0000} \textbf{9}}  & {\color[HTML]{FE0000} \textbf{10}} & {\color[HTML]{FE0000} \textbf{9}}  & {\color[HTML]{FE0000} \textbf{11}} & {\color[HTML]{FE0000} \textbf{10}} & {\color[HTML]{FE0000} \textbf{10}} & {\color[HTML]{FE0000} \textbf{10}} & {\color[HTML]{FE0000} \textbf{9}}  & \textbf{15425}                     & {\color[HTML]{FE0000} \textbf{10}} & {\color[HTML]{FE0000} \textbf{10}} & {\color[HTML]{FE0000} \textbf{10}} & {\color[HTML]{FE0000} \textbf{12}} \\ 
    {\color[HTML]{9A0000} \textbf{LT}}                                                                     & {\color[HTML]{FE0000} \textbf{10}} & 1036                               & 1193                               & 1423                               & 1624                               & 1726                               & 1555                               & 1332                               & {\color[HTML]{FE0000} \textbf{10}} & \textbf{3261}                      & 306                                & 870                                & 1430                               \\ 
    {\color[HTML]{9A0000} \textbf{IN}}                                                                     & {\color[HTML]{FE0000} \textbf{10}} & 206                                & 241                                & 298                                & 343                                & 390                                & 372                                & 217                                & {\color[HTML]{FE0000} \textbf{10}} & 327                                & \textbf{733}                       & 286                                & 349                                \\ 
    {\color[HTML]{9A0000} \textbf{UN}}                                                                     & {\color[HTML]{FE0000} \textbf{11}} & 973                                & 1317                               & 1859                               & 2116                               & 2299                               & 1997                               & 1343                               & {\color[HTML]{FE0000} \textbf{10}} & 1766                               & 704                                & \textbf{1927}                      & 2010                               \\ 
    {\color[HTML]{9A0000} \textbf{DN}}                                                                     & {\color[HTML]{FE0000} \textbf{10}} & 811                                & 973                                & 1218                               & 1399                               & 1513                               & 1171                               & 1018                               & {\color[HTML]{FE0000} \textbf{10}} & 1265                               & 347                                & 821                                & \textbf{1444}                      \\ 
    \multicolumn{1}{l}{}                                                                                 & \multicolumn{13}{c}{{\color[HTML]{3531FF} \textbf{Patch size = 512 x 512 pixels ($\sim$3000 patches per camera per pipeline)}}}                                                                                                                                                                                                                                                                                                                                                                \\ \midrule
    {\color[HTML]{9A0000} \textbf{RT}}                                                                     & \textbf{44}                        & {\color[HTML]{FE0000} \textbf{10}} & {\color[HTML]{FE0000} \textbf{10}} & {\color[HTML]{FE0000} \textbf{11}} & {\color[HTML]{FE0000} \textbf{11}} & {\color[HTML]{FE0000} \textbf{11}} & {\color[HTML]{FE0000} \textbf{10}} & {\color[HTML]{FE0000} \textbf{10}} & {\color[HTML]{FE0000} \textbf{10}} & {\color[HTML]{FE0000} \textbf{10}} & {\color[HTML]{FE0000} \textbf{11}} & {\color[HTML]{FE0000} \textbf{11}} & {\color[HTML]{FE0000} \textbf{12}} \\ 
    {\color[HTML]{9A0000} \textbf{AT}}                                                                     & {\color[HTML]{FE0000} \textbf{10}} & \textbf{200}                       & 203                                & 196                                & 221                                & 226                                & 169                                & 218                                & {\color[HTML]{FE0000} \textbf{10}} & 243                                & 44                                 & 111                                & 176                                \\ 
    {\color[HTML]{9A0000} \textbf{LZ}}                                                                     & {\color[HTML]{FE0000} \textbf{11}} & 427                                & \textbf{554}                       & 538                                & 626                                & 650                                & 517                                & 519                                & {\color[HTML]{FE0000} \textbf{10}} & 590                                & 135                                & 334                                & 500                                \\ 
    {\color[HTML]{9A0000} \textbf{DR}}                                                                     & {\color[HTML]{FE0000} \textbf{11}} & 379                                & 458                                & \textbf{653}                       & 704                                & 747                                & 479                                & 446                                & {\color[HTML]{FE0000} \textbf{10}} & 525                                & 143                                & 386                                & 557                                \\ 
    {\color[HTML]{9A0000} \textbf{LR}}                                                                     & {\color[HTML]{FE0000} \textbf{11}} & 421                                & 527                                & 699                                & \textbf{811}                       & 856                                & 556                                & 518                                & {\color[HTML]{FE0000} \textbf{10}} & 609                                & 169                                & 437                                & 639                                \\ 
    {\color[HTML]{9A0000} \textbf{AF}}                                                                     & {\color[HTML]{FE0000} \textbf{11}} & 437                                & 546                                & 735                                & 846                                & \textbf{906}                       & 590                                & 541                                & {\color[HTML]{FE0000} \textbf{10}} & 637                                & 187                                & 471                                & 661                                \\ 
    {\color[HTML]{9A0000} \textbf{LM}}                                                                     & {\color[HTML]{FE0000} \textbf{11}} & 480                                & 682                                & 757                                & 872                                & 930                                & \textbf{1201}                      & 718                                & {\color[HTML]{FE0000} \textbf{10}} & 920                                & 245                                & 601                                & 735                                \\ 
    {\color[HTML]{9A0000} \textbf{DX}}                                                                     & {\color[HTML]{FE0000} \textbf{10}} & 215                                & 247                                & 239                                & 275                                & 286                                & 250                                & \textbf{272}                       & {\color[HTML]{FE0000} \textbf{10}} & 327                                & 47                                 & 142                                & 224                                \\ 
    {\color[HTML]{9A0000} \textbf{C1}}                                                                     & {\color[HTML]{FE0000} \textbf{10}} & {\color[HTML]{FE0000} \textbf{10}} & {\color[HTML]{FE0000} \textbf{10}} & {\color[HTML]{FE0000} \textbf{11}} & {\color[HTML]{FE0000} \textbf{10}} & {\color[HTML]{FE0000} \textbf{10}} & {\color[HTML]{FE0000} \textbf{10}} & {\color[HTML]{FE0000} \textbf{9}}  & \textbf{3374}                      & {\color[HTML]{FE0000} \textbf{10}} & {\color[HTML]{FE0000} \textbf{10}} & {\color[HTML]{FE0000} \textbf{11}} & {\color[HTML]{FE0000} \textbf{11}} \\ 
    {\color[HTML]{9A0000} \textbf{LT}}                                                                     & {\color[HTML]{FE0000} \textbf{10}} & 243                                & 288                                & 289                                & 332                                & 346                                & 326                                & 313                                & {\color[HTML]{FE0000} \textbf{10}} & \textbf{681}                       & 67                                 & 182                                & 272                                \\ 
    {\color[HTML]{9A0000} \textbf{IN}}                                                                     & {\color[HTML]{FE0000} \textbf{10}} & 45                                 & 57                                 & 63                                 & 69                                 & 74                                 & 71                                 & 46                                 & {\color[HTML]{FE0000} \textbf{10}} & 68                                 & \textbf{142}                       & 58                                 & 65                                 \\ 
    {\color[HTML]{9A0000} \textbf{UN}}                                                                     & {\color[HTML]{FE0000} \textbf{11}} & 242                                & 309                                & 429                                & 478                                & 510                                & 421                                & 299                                & {\color[HTML]{FE0000} \textbf{10}} & 373                                & 158                                & \textbf{454}                       & 447                                \\ 
    {\color[HTML]{9A0000} \textbf{DN}}                                                                     & {\color[HTML]{FE0000} \textbf{10}} & 176                                & 216                                & 248                                & 278                                & 295                                & 229                                & 215                                & {\color[HTML]{FE0000} \textbf{10}} & 264                                & 70                                 & 166                                & \textbf{279}                       \\ 
    \multicolumn{1}{l}{}                                                                                 & \multicolumn{13}{c}{{\color[HTML]{3531FF} \textbf{Patch size = 256 x 256 pixels ($\sim$13000 patches per camera per pipeline)}}}                                                                                                                                                                                                                                                                                                                                                               \\ \midrule
    {\color[HTML]{9A0000} \textbf{RT}}                                                                     & \textbf{15}                        & {\color[HTML]{FE0000} \textbf{10}} & {\color[HTML]{FE0000} \textbf{10}} & {\color[HTML]{FE0000} \textbf{11}} & {\color[HTML]{FE0000} \textbf{11}} & {\color[HTML]{FE0000} \textbf{11}} & {\color[HTML]{FE0000} \textbf{10}} & {\color[HTML]{FE0000} \textbf{10}} & {\color[HTML]{FE0000} \textbf{10}} & {\color[HTML]{FE0000} \textbf{10}} & {\color[HTML]{FE0000} \textbf{11}} & {\color[HTML]{FE0000} \textbf{11}} & {\color[HTML]{FE0000} \textbf{12}} \\ 
    {\color[HTML]{9A0000} \textbf{AT}}                                                                     & {\color[HTML]{FE0000} \textbf{10}} & \textbf{54}                        & 52                                 & 47                                 & 52                                 & 53                                 & 44                                 & 58                                 & {\color[HTML]{FE0000} \textbf{10}} & 62                                 & 15                                 & 27                                 & 41                                 \\ 
    {\color[HTML]{9A0000} \textbf{LZ}}                                                                     & {\color[HTML]{FE0000} \textbf{11}} & 111                                & \textbf{153}                       & 132                                & 154                                & 158                                & 131                                & 135                                & {\color[HTML]{FE0000} \textbf{10}} & 151                                & 36                                 & 83                                 & 114                                \\ 
    {\color[HTML]{9A0000} \textbf{DR}}                                                                     & {\color[HTML]{FE0000} \textbf{11}} & 90                                 & 114                                & \textbf{161}                       & 172                                & 178                                & 109                                & 106                                & {\color[HTML]{FE0000} \textbf{10}} & 121                                & 37                                 & 94                                 & 129                                \\ 
    {\color[HTML]{9A0000} \textbf{LR}}                                                                     & {\color[HTML]{FE0000} \textbf{11}} & 103                                & 132                                & 174                                & \textbf{201}                       & 207                                & 128                                & 124                                & {\color[HTML]{FE0000} \textbf{10}} & 141                                & 44                                 & 111                                & 150                                \\ 
    {\color[HTML]{9A0000} \textbf{AF}}                                                                     & {\color[HTML]{FE0000} \textbf{11}} & 105                                & 137                                & 180                                & 205                                & \textbf{218}                       & 135                                & 127                                & {\color[HTML]{FE0000} \textbf{10}} & 148                                & 48                                 & 118                                & 156                                \\ 
    {\color[HTML]{9A0000} \textbf{LM}}                                                                     & {\color[HTML]{FE0000} \textbf{11}} & 126                                & 174                                & 173                                & 203                                & 209                                & \textbf{309}                       & 183                                & {\color[HTML]{FE0000} \textbf{10}} & 232                                & 58                                 & 138                                & 160                                \\ 
    {\color[HTML]{9A0000} \textbf{DX}}                                                                     & {\color[HTML]{FE0000} \textbf{10}} & 57                                 & 63                                 & 56                                 & 64                                 & 65                                 & 65                                 & \textbf{72}                        & {\color[HTML]{FE0000} \textbf{10}} & 85                                 & 15                                 & 34                                 & 50                                 \\ 
    {\color[HTML]{9A0000} \textbf{C1}}                                                                     & {\color[HTML]{FE0000} \textbf{10}} & {\color[HTML]{FE0000} \textbf{10}} & {\color[HTML]{FE0000} \textbf{10}} & {\color[HTML]{FE0000} \textbf{11}} & {\color[HTML]{FE0000} \textbf{10}} & {\color[HTML]{FE0000} \textbf{10}} & {\color[HTML]{FE0000} \textbf{10}} & {\color[HTML]{FE0000} \textbf{10}} & \textbf{932}                       & {\color[HTML]{FE0000} \textbf{10}} & {\color[HTML]{FE0000} \textbf{10}} & {\color[HTML]{FE0000} \textbf{10}} & {\color[HTML]{FE0000} \textbf{11}} \\ 
    {\color[HTML]{9A0000} \textbf{LT}}                                                                     & {\color[HTML]{FE0000} \textbf{10}} & 64                                 & 73                                 & 67                                 & 77                                 & 79                                 & 83                                 & 82                                 & {\color[HTML]{FE0000} \textbf{10}} & \textbf{178}                       & 18                                 & 43                                 & 61                                 \\ 
    {\color[HTML]{9A0000} \textbf{IN}}                                                                     & {\color[HTML]{FE0000} \textbf{10}} & 14                                 & 16                                 & 19                                 & 19                                 & 20                                 & 18                                 & 14                                 & {\color[HTML]{FE0000} \textbf{10}} & 18                                 & \textbf{37}                        & 17                                 & 19                                 \\ 
    {\color[HTML]{9A0000} \textbf{UN}}                                                                     & {\color[HTML]{FE0000} \textbf{11}} & 58                                 & 77                                 & 108                                & 117                                & 123                                & 97                                 & 71                                 & {\color[HTML]{FE0000} \textbf{10}} & 86                                 & 41                                 & \textbf{110}                       & 103                                \\ 
    {\color[HTML]{9A0000} \textbf{DN}}                                                                     & {\color[HTML]{FE0000} \textbf{10}} & 43                                 & 54                                 & 63                                 & 70                                 & 73                                 & 54                                 & 52                                 & {\color[HTML]{FE0000} \textbf{10}} & 64                                 & 20                                 & 42                                 & \textbf{67}                        \\ 
    \textbf{}                                                                                              & \multicolumn{13}{c}{{\color[HTML]{3531FF} \textbf{Patch size = 128 x 128 pixels ($\sim$55000 patches per camera per pipeline)}}}                                                                                                                                                                                                                                                                                                                                                              \\ \midrule
    {\color[HTML]{9A0000} \textbf{RT}}                                                                     & \textbf{11}                        & {\color[HTML]{FE0000} \textbf{10}} & {\color[HTML]{FE0000} \textbf{10}} & {\color[HTML]{FE0000} \textbf{11}} & {\color[HTML]{FE0000} \textbf{11}} & {\color[HTML]{FE0000} \textbf{11}} & {\color[HTML]{FE0000} \textbf{10}} & {\color[HTML]{FE0000} \textbf{10}} & {\color[HTML]{FE0000} \textbf{10}} & {\color[HTML]{FE0000} \textbf{10}} & {\color[HTML]{FE0000} \textbf{10}} & {\color[HTML]{FE0000} \textbf{11}} & {\color[HTML]{FE0000} \textbf{11}} \\ 
    {\color[HTML]{9A0000} \textbf{AT}}                                                                     & {\color[HTML]{FE0000} \textbf{10}} & \textbf{17}                        & 15                                 & 14                                 & 15                                 & 15                                 & 14                                 & 17                                 & {\color[HTML]{FE0000} \textbf{10}} & 17                                 & 11                                 & 13                                 & 14                                 \\ 
    {\color[HTML]{9A0000} \textbf{LZ}}                                                                     & {\color[HTML]{FE0000} \textbf{11}} & 28                                 & \textbf{40}                        & 30                                 & 35                                 & 35                                 & 31                                 & 34                                 & {\color[HTML]{FE0000} \textbf{10}} & 37                                 & 14                                 & 21                                 & 26                                 \\ 
    {\color[HTML]{9A0000} \textbf{DR}}                                                                     & {\color[HTML]{FE0000} \textbf{11}} & 21                                 & 26                                 & \textbf{38}                        & 40                                 & 42                                 & 24                                 & 24                                 & {\color[HTML]{FE0000} \textbf{10}} & 27                                 & 14                                 & 24                                 & 30                                 \\ 
    {\color[HTML]{9A0000} \textbf{LR}}                                                                     & {\color[HTML]{FE0000} \textbf{11}} & 24                                 & 31                                 & 42                                 & \textbf{48}                        & 50                                 & 28                                 & 29                                 & {\color[HTML]{FE0000} \textbf{10}} & 32                                 & 15                                 & 28                                 & 35                                 \\ 
    {\color[HTML]{9A0000} \textbf{AF}}                                                                     & {\color[HTML]{FE0000} \textbf{11}} & 24                                 & 32                                 & 43                                 & 49                                 & \textbf{52}                        & 29                                 & 29                                 & {\color[HTML]{FE0000} \textbf{10}} & 33                                 & 15                                 & 29                                 & 36                                 \\ 
    {\color[HTML]{9A0000} \textbf{LM}}                                                                     & {\color[HTML]{FE0000} \textbf{11}} & 31                                 & 41                                 & 36                                 & 42                                 & 43                                 & \textbf{75}                        & 45                                 & {\color[HTML]{FE0000} \textbf{10}} & 57                                 & 16                                 & 29                                 & 32                                 \\ 
    {\color[HTML]{9A0000} \textbf{DX}}                                                                     & {\color[HTML]{FE0000} \textbf{10}} & 17                                 & 17                                 & 16                                 & 16                                 & 16                                 & 17                                 & \textbf{20}                        & {\color[HTML]{FE0000} \textbf{10}} & 22                                 & 11                                 & 13                                 & 15                                 \\ 
    {\color[HTML]{9A0000} \textbf{C1}}                                                                     & {\color[HTML]{FE0000} \textbf{10}} & {\color[HTML]{FE0000} \textbf{10}} & {\color[HTML]{FE0000} \textbf{10}} & {\color[HTML]{FE0000} \textbf{10}} & {\color[HTML]{FE0000} \textbf{10}} & {\color[HTML]{FE0000} \textbf{10}} & {\color[HTML]{FE0000} \textbf{10}} & {\color[HTML]{FE0000} \textbf{10}} & \textbf{240}                       & {\color[HTML]{FE0000} \textbf{10}} & {\color[HTML]{FE0000} \textbf{10}} & {\color[HTML]{FE0000} \textbf{10}} & {\color[HTML]{FE0000} \textbf{11}} \\ 
    {\color[HTML]{9A0000} \textbf{LT}}                                                                     & {\color[HTML]{FE0000} \textbf{11}} & 18                                 & 19                                 & 17                                 & 18                                 & 18                                 & 21                                 & 22                                 & {\color[HTML]{FE0000} \textbf{10}} & \textbf{46}                        & 12                                 & 14                                 & 16                                 \\ 
    {\color[HTML]{9A0000} \textbf{IN}}                                                                     & {\color[HTML]{FE0000} \textbf{10}} & 11                                 & 11                                 & 12                                 & 12                                 & 12                                 & 11                                 & 11                                 & {\color[HTML]{FE0000} \textbf{10}} & 11                                 & \textbf{14}                        & 12                                 & 12                                 \\ 
    {\color[HTML]{9A0000} \textbf{UN}}                                                                     & {\color[HTML]{FE0000} \textbf{11}} & 16                                 & 19                                 & 26                                 & 29                                 & 30                                 & 21                                 & 18                                 & {\color[HTML]{FE0000} \textbf{10}} & 20                                 & 15                                 & \textbf{28}                        & 25                                 \\ 
    {\color[HTML]{9A0000} \textbf{DN}}                                                                     & {\color[HTML]{FE0000} \textbf{10}} & 14                                 & 15                                 & 18                                 & 19                                 & 19                                 & 15                                 & 15                                 & {\color[HTML]{FE0000} \textbf{10}} & 16                                 & 12                                 & 14                                 & \textbf{18}                        \\ \bottomrule
    \end{tabular}
    }
    \end{table}

\subsection{Impact on ROC Curves and Error Rates}
\label{subsec:ISP_ROC}

To assess the impact of ISP mismatch on effective error rates, we generate full receiver operating characteristics (ROC curves). 
An example set of ROC curves for smaller patch sizes (i.e., 256 and 128 px), selective cameras, and a single PRNU estimation ISP (LibRaw) is shown in Figure~\ref{fig:ISP_ROC_PatchSizes}.
Analogously to previous experiments, we can see a significant variation of matching performance with the strongest deterioration for small patch sizes and for neural ISPs. 

To quantitatively summarize the degradation, we show true positive rates corresponding to a fixed 0.5\% false-positive rate (Table~\ref{tab:ISP_TPR_LibRaw_256_128}). We focus on 128 and 256~px patches due to the much larger number of available samples. For the illustrated LibRaw and Affinity Photo ISPs, the average deterioration in TPR was 17 percentage points (13~ppt for well-synchronized standard darkroom software and 20~ppt for neural ISPs).

\begin{table}[htb!]
    \centering
    \caption{TPR at 0.5 \% FPR using PRNU pipe (LibRaw and Affinity)  and test images (of all pipelines) for cross-pipeline experiments}
    \label{tab:ISP_TPR_LibRaw_256_128}
    \resizebox{0.49\textwidth}{!}{
    \begin{tabular}{cccccccccccccc}
    \toprule
    {\color[HTML]{9A0000} }                                  & \multicolumn{13}{c}{{\color[HTML]{3531FF} \textbf{Camera Processing Pipelines for Test Images}}}                                                                                                                                                                                                                                                  \\ \cline{2-14} \\
    \multirow{-2}{*}{{\color[HTML]{9A0000} \textbf{Camera}}} & {\color[HTML]{3531FF} \textbf{RT}} & {\color[HTML]{3531FF} \textbf{AT}} & {\color[HTML]{3531FF} \textbf{LZ}} & {\color[HTML]{3531FF} \textbf{DR}} & {\color[HTML]{3531FF} \textbf{LR}} & {\color[HTML]{3531FF} \textbf{AF}} & {\color[HTML]{3531FF} \textbf{LM}} & {\color[HTML]{3531FF} \textbf{DX}} & {\color[HTML]{3531FF} \textbf{C1}} & {\color[HTML]{3531FF} \textbf{LT}} & {\color[HTML]{3531FF} \textbf{IN}} & {\color[HTML]{3531FF} \textbf{UN}} & {\color[HTML]{3531FF} \textbf{DN}} \\ \midrule
    {\color[HTML]{9A0000} }                                  & \multicolumn{13}{c}{{\color[HTML]{3531FF} \textbf{\begin{tabular}[c]{@{}c@{}}PRNU pipe = LibRaw\\ Patch size = 256 x 256 pixels ($\sim$40,000 patches per camera per pipeline)\end{tabular}}}}                                                                                                                                                                                                                                                  \\ \cline{2-14} 
    {\color[HTML]{9A0000} \textbf{N7k}}                      & 0.25                               & 0.88                               & 0.91                               & 0.93                               & \textbf{0.95}                      & 0.94                               & 0.92                               & 0.90                               & 0.07                               & 0.94                               & 0.69                               & 0.85                               & 0.88                               \\ 
    {\color[HTML]{9A0000} \textbf{N90}}                      & 0.06                               & 0.68                               & 0.75                               & 0.73                               & \textbf{0.77}                      & 0.76                               & 0.69                               & 0.73                               & 0.06                               & 0.72                               & 0.47                               & 0.68                               & 0.58                               \\ 
    {\color[HTML]{9A0000} \textbf{C4D}}                      & 0.72                               & -                                  & 0.72                               & 0.73                               & \textbf{0.75}                      & 0.75                               & 0.70                               & 0.73                               & 0.67                               & 0.73                               & 0.44                               & 0.22                               & 0.62                               \\ 
    {\color[HTML]{9A0000} \textbf{C5D}}                      & 0.69                               & -                                  & 0.69                               & 0.68                               & \textbf{0.69}                      & 0.70                               & 0.64                               & 0.68                               & 0.63                               & 0.67                               & 0.47                               & 0.57                               & 0.59                               \\ 
    \multicolumn{1}{l}{}                                   & \multicolumn{13}{c}{{\color[HTML]{3531FF} \textbf{Patch size = 128 x 128 pixels ($\sim$155,000 patches per pipeline)}}}                                                                                                                                                                                                                                                                                                                                                             \\ \midrule
    {\color[HTML]{9A0000} \textbf{N7k}}                      & 0.23                               & 0.59                               & 0.66                               & 0.69                               & \textbf{0.74}                      & 0.75                               & 0.64                               & 0.63                               & 0.06                               & 0.68                               & 0.44                               & 0.61                               & 0.65                               \\ 
    {\color[HTML]{9A0000} \textbf{N90}}                      & 0.07                               & 0.47                               & 0.54                               & 0.54                               & \textbf{0.58}                      & 0.58                               & 0.46                               & 0.49                               & 0.06                               & 0.48                               & 0.26                               & 0.48                               & 0.41                               \\ 
    {\color[HTML]{9A0000} \textbf{C4D}}                      & 0.45                               & -                                  & 0.43                               & 0.49                               & \textbf{0.50}                      & 0.51                               & 0.43                               & 0.45                               & 0.38                               & 0.42                               & 0.23                               & 0.10                               & 0.40                               \\ 
    {\color[HTML]{9A0000} \textbf{C5D}}                      & 0.51                               & -                                  & 0.51                               & 0.51                               & \textbf{0.54}                      & 0.55                               & 0.44                               & 0.49                               & 0.42                               & 0.45                               & 0.28                               & 0.42                               & 0.43                               \\ \bottomrule \vspace{0pt} \\
    \multicolumn{1}{l}{}                                   & \multicolumn{13}{c}{{\color[HTML]{3531FF} \textbf{\begin{tabular}[c]{@{}c@{}}PRNU pipe = Affinity\\ Patch size = 256 x 256 pixels ($\sim$40,000 patches per camera per pipeline)\end{tabular}}}}                                                                                                                                                                                                                                                                                              \\ \midrule
    {\color[HTML]{9A0000} \textbf{N7k}}                      & \multicolumn{1}{l}{0.26}          & \multicolumn{1}{l}{0.89}          & \multicolumn{1}{l}{0.92}          & \multicolumn{1}{l}{0.93}          & \multicolumn{1}{l}{0.95}          & \multicolumn{1}{l}{\textbf{0.94}} & \multicolumn{1}{l}{0.92}          & \multicolumn{1}{l}{0.90}          & \multicolumn{1}{l}{0.07}          & \multicolumn{1}{l}{0.94}          & \multicolumn{1}{l}{0.70}          & \multicolumn{1}{l}{0.86}          & \multicolumn{1}{l}{0.88}          \\ 
    {\color[HTML]{9A0000} \textbf{N90}}                      & \multicolumn{1}{l}{0.07}          & \multicolumn{1}{l}{0.70}          & \multicolumn{1}{l}{0.76}          & \multicolumn{1}{l}{0.74}          & \multicolumn{1}{l}{0.78}          & \multicolumn{1}{l}{\textbf{0.78}} & \multicolumn{1}{l}{0.70}          & \multicolumn{1}{l}{0.73}          & \multicolumn{1}{l}{0.06}          & \multicolumn{1}{l}{0.73}          & \multicolumn{1}{l}{0.48}          & \multicolumn{1}{l}{0.68}          & \multicolumn{1}{l}{0.59}          \\ 
    {\color[HTML]{9A0000} \textbf{C4D}}                      & \multicolumn{1}{l}{0.73}          & \multicolumn{1}{l}{-}             & \multicolumn{1}{l}{0.72}          & \multicolumn{1}{l}{0.74}          & \multicolumn{1}{l}{0.76}          & \multicolumn{1}{l}{\textbf{0.76}} & \multicolumn{1}{l}{0.71}          & \multicolumn{1}{l}{0.74}          & \multicolumn{1}{l}{0.68}          & \multicolumn{1}{l}{0.73}          & \multicolumn{1}{l}{0.45}          & \multicolumn{1}{l}{0.24}          & \multicolumn{1}{l}{0.64}          \\ 
    {\color[HTML]{9A0000} \textbf{C5D}}                      & \multicolumn{1}{l}{0.68}          & \multicolumn{1}{l}{-}             & \multicolumn{1}{l}{0.68}          & \multicolumn{1}{l}{0.68}          & \multicolumn{1}{l}{0.69}          & \multicolumn{1}{l}{\textbf{0.70}} & \multicolumn{1}{l}{0.64}          & \multicolumn{1}{l}{0.68}          & \multicolumn{1}{l}{0.63}          & \multicolumn{1}{l}{0.66}          & \multicolumn{1}{l}{0.46}          & \multicolumn{1}{l}{0.57}          & \multicolumn{1}{l}{0.59}          \\ 
                                                             & \multicolumn{13}{c}{{\color[HTML]{3531FF} \textbf{Patch size = 128 x 128 pixels ($\sim$1,55,000 patches per pipeline)}}}                                                                                                                                                                                                                                                                                                                                                             \\ \midrule
    {\color[HTML]{9A0000} \textbf{N7k}}                      & 0.21                               & 0.60                               & 0.67                               & 0.70                               & 0.75                               & \textbf{0.76}                      & 0.65                               & 0.63                               & 0.06                               & 0.69                               & 0.46                               & 0.62                               & 0.66                               \\ 
    {\color[HTML]{9A0000} \textbf{N90}}                      & 0.07                               & 0.49                               & 0.55                               & 0.55                               & 0.58                               & \textbf{0.59}                      & 0.48                               & 0.51                               & 0.06                               & 0.50                               & 0.28                               & 0.50                               & 0.43                               \\ 
    {\color[HTML]{9A0000} \textbf{C4D}}                      & 0.46                               & -                                  & 0.45                               & 0.50                               & 0.52                               & \textbf{0.53}                      & 0.45                               & 0.47                               & 0.40                               & 0.44                               & 0.23                               & 0.11                               & 0.42                               \\ 
    {\color[HTML]{9A0000} \textbf{C5D}}                      & 0.49                               & -                                  & 0.49                               & 0.50                               & 0.52                               & \textbf{0.54}                      & 0.42                               & 0.47                               & 0.40                               & 0.43                               & 0.25                               & 0.41                               & 0.43                               \\ \bottomrule
    \end{tabular}
    }
    \end{table}

\section{Conclusion and Future Work}
\label{sec:ISP_Conclusion}

% {\color{red}I haven't revised this yet. Needs some reworking...}

% In the absence of off-the-shelf camera processing pipelines, a camera fingerprint does not have multiple versions. 
% Whereas, given the existence of such pipelines, we needed to compute multiple PRNU versions (i.e., thirteen), one for each available pipeline.

Our results illustrate that camera fingerprints exhibit non-negligible variation with the imaging pipeline. Despite choosing the most conservative setting with limited enhancement (e.g., default settings in darkroom software) and fixed selection of images for PRNU estimation, we observed significant deterioration in all metrics, starting with fingerprint correlation, up to detection statistics and real-world performance metrics. This means that forensic analysts should be aware of a potential ISP mismatch, especially when using small analysis windows (e.g., manipulation localization) or when stronger enhancement or post-processing may be involved.

We experimented with 13 different pipelines representing various digital darkrooms and neural networks. The degradation was the strongest for smaller patches which are commonly used in photo manipulation detection. Specifically, for false positive rate fixed at 0.5\%, an average deterioration of 17 percentage points was observed for patches of size 128~px (sample size 55,000) and 256~px (sample size 13,000). To illustrate qualitative impact on tampering localization, we show an example forgery and the corresponding authentication results in Figure~\ref{fig:ISP_localization}. The depicted example is a synthetic forgery created by pasting a foreign patch of 1024~px inside a full resolution image taken with Nikon~7000 and processed by each pipeline. We show both local PCE responses and tampering probabilities obtained from \emph{p}-values of PCE statistics~\cite{goljan2008digital}. It can be clearly observed that even for conventional darkroom software, the results exhibit strong variation.

We expect that the diversity of RAW processing pipelines, and increasing adoption of computational photography and machine learning (also directly inside of the cameras) will have more significant impact on PRNU analysis in the future. Negative impact of this variation is likely to be most pronounced not only in tampering localization, but also in reduced forms of PRNU used in large-scale search and  attribution~\cite{bayram2014sensor,valsesia2015compressed,bondi2018improving,taspinar2017fast}. 
We have run exhaustive experiments, but in a limited, well-controlled lab setting. In most recent evaluations, other researchers are also starting to question fingerprint uniqueness in modern smartphones with advanced cameras~\cite{iuliani2020leak}. Further work will be needed to assess the scope of the problem and to design new solutions. Some early work in this direction includes PRNU analysis in high-dynamic-range (HDR) photographs \cite{darvish2019camera}.

\section*{Acknowledgement}
\label{sec:ISP_Ack}
This material is based upon work partially supported by  the  overseas research experience fellowship, Indian Institute of Technology Gandhinagar (IITGN), the Department of Science and Technology (DST), Government of India under the Award Number ECR/2015/000583, and supported by Visvesvaraya Ph.D. Scheme, Ministry of the Electronics and Information Technology (MeitY), Government of India MEITY-PHD-951.
Any opinions, findings, and conclusions or recommendations expressed in this material are those of the author(s) and do not necessarily reflect the views of the funding agencies.

% % use section* for acknowledgment
% \section*{Acknowledgment}

\bibliographystyle{IEEEtran}
\bibliography{ISP}

\newpage
% \section*{Supplementary Material}
% \label{sec:ISP_Supp}
\begin{table*}[b!]
    \centering
    \caption{Median PCE values for cross-pipeline attribution of non-overlapping patches from Canon EOS 40D}
    \label{tab:ISP_PCE_Med_C4}
    % \resizebox{0.49\textwidth}{!}{
    \begin{tabular}{cccccccccccccc}
    \toprule
    {\color[HTML]{9A0000} }                                                                                & \multicolumn{13}{c}{{\color[HTML]{3531FF} \textbf{Image Processing Pipelines for Test Images}}}                                                                                                                                                                                                                                                                       \\ \cline{2-14} \\
    \multirow{-2}{*}{{\color[HTML]{9A0000} \textbf{\begin{tabular}[c]{@{}c@{}}PRNU \\ ISP\end{tabular}}}} & {\color[HTML]{3531FF} \textbf{RT}} & {\color[HTML]{3531FF} \textbf{AT}} & {\color[HTML]{3531FF} \textbf{LZ}} & {\color[HTML]{3531FF} \textbf{DR}} & {\color[HTML]{3531FF} \textbf{LR}} & {\color[HTML]{3531FF} \textbf{AF}} & {\color[HTML]{3531FF} \textbf{LM}} & {\color[HTML]{3531FF} \textbf{DX}} & {\color[HTML]{3531FF} \textbf{C1}} & {\color[HTML]{3531FF} \textbf{LT}} & {\color[HTML]{3531FF} \textbf{IN}} & {\color[HTML]{3531FF} \textbf{UN}} & {\color[HTML]{3531FF} \textbf{DN}} \\ \midrule
    {\color[HTML]{9A0000} }                                                                                & \multicolumn{13}{c}{{\color[HTML]{3531FF} \textbf{Patch size = 1024 x 1024 pixels ($\sim$600 patches per camera per pipeline)}}}                                                                                                                                                                                                                                                                       \\ \midrule
    
    {\color[HTML]{9A0000} \textbf{RT}}                                                                     & \textbf{966}                       & {\color[HTML]{333333} -}           & {\color[HTML]{333333} 758}         & {\color[HTML]{333333} 888}         & {\color[HTML]{333333} 1056}        & {\color[HTML]{333333} 1163}        & {\color[HTML]{333333} 1122}        & {\color[HTML]{333333} 966}         & {\color[HTML]{333333} 786}         & {\color[HTML]{333333} 825}         & {\color[HTML]{333333} 146}         & {\color[HTML]{333333} 26}          & {\color[HTML]{333333} 702}         \\
{\color[HTML]{9A0000} \textbf{AT}}                                                                     & {\color[HTML]{333333} -}           & {\color[HTML]{333333} -}           & {\color[HTML]{333333} -}           & {\color[HTML]{333333} -}           & {\color[HTML]{333333} -}           & {\color[HTML]{333333} -}           & {\color[HTML]{333333} -}           & {\color[HTML]{333333} -}           & {\color[HTML]{333333} -}           & {\color[HTML]{333333} -}           & {\color[HTML]{333333} -}           & {\color[HTML]{333333} -}           & {\color[HTML]{333333} -}           \\
{\color[HTML]{9A0000} \textbf{LZ}}                                                                     & {\color[HTML]{333333} 869}         & -                                  & \textbf{825}                       & 949                                & 1124                               & 1094                               & 1018                               & 912                                & {\color[HTML]{333333} 719}         & 828                                & 161                                & 26                                 & 871                                \\
{\color[HTML]{9A0000} \textbf{DR}}                                                                     & {\color[HTML]{333333} 688}         & -                                  & 610                                & \textbf{763}                       & 873                                & 957                                & 840                                & 696                                & {\color[HTML]{333333} 607}         & 641                                & 106                                & 24                                 & 607                                \\
{\color[HTML]{9A0000} \textbf{LR}}                                                                     & {\color[HTML]{333333} 960}         & -                                  & 856                                & 1168                               & \textbf{1319}                      & 1359                               & 1202                               & 972                                & {\color[HTML]{333333} 847}         & 891                                & 172                                & 28                                 & 933                                \\
{\color[HTML]{9A0000} \textbf{AF}}                                                                     & {\color[HTML]{333333} 1044}        & -                                  & 866                                & 1122                               & 1330                               & \textbf{1508}                      & 1274                               & 1072                               & {\color[HTML]{333333} 876}         & 966                                & 158                                & 31                                 & 939                                \\
{\color[HTML]{9A0000} \textbf{LM}}                                                                     & {\color[HTML]{333333} 1166}        & -                                  & 948                                & 1257                               & 1476                               & 1518                               & \textbf{1798}                      & 1120                               & {\color[HTML]{333333} 1086}        & 1064                               & 125                                & 35                                 & 1007                               \\
{\color[HTML]{9A0000} \textbf{DX}}                                                                     & {\color[HTML]{333333} 904}         & -                                  & 783                                & 958                                & 1066                               & 1094                               & 1046                               & \textbf{1048}                      & {\color[HTML]{333333} 767}         & 854                                & 164                                & 26                                 & 853                                \\
{\color[HTML]{9A0000} \textbf{C1}}                                                                     & {\color[HTML]{333333} 460}         & {\color[HTML]{333333} -}           & {\color[HTML]{333333} 366}         & {\color[HTML]{333333} 493}         & {\color[HTML]{333333} 570}         & {\color[HTML]{333333} 575}         & {\color[HTML]{333333} 650}         & {\color[HTML]{333333} 473}         & \textbf{18923}                     & {\color[HTML]{333333} 370}         & {\color[HTML]{333333} 66}          & {\color[HTML]{333333} 21}          & {\color[HTML]{333333} 403}         \\
{\color[HTML]{9A0000} \textbf{LT}}                                                                     & {\color[HTML]{333333} 849}         & -                                  & 777                                & 953                                & 1093                               & 1085                               & 1086                               & 938                                & {\color[HTML]{333333} 722}         & \textbf{3095}                      & 145                                & 26                                 & 817                                \\
{\color[HTML]{9A0000} \textbf{IN}}                                                                     & {\color[HTML]{333333} 137}         & -                                  & 126                                & 144                                & 157                                & 164                                & 109                                & 156                                & {\color[HTML]{333333} 102}         & 126                                & \textbf{1301}                      & 19                                 & 136                                \\
{\color[HTML]{9A0000} \textbf{UN}}                                                                     & {\color[HTML]{333333} 20}          & -                                  & 19                                 & 29                                 & 29                                 & 26                                 & 27                                 & 19                                 & {\color[HTML]{333333} 20}          & 20                                 & 17                                 & \textbf{75}                        & 29                                 \\
{\color[HTML]{9A0000} \textbf{DN}}                                                                     & {\color[HTML]{333333} 715}         & -                                  & 633                                & 853                                & 940                                & 1013                               & 874                                & 712                                & {\color[HTML]{333333} 590}         & 681                                & 137                                & 25                                 & \textbf{964}                       \\ 
 
    \multicolumn{1}{l}{}                                                                                 & \multicolumn{13}{c}{{\color[HTML]{3531FF} \textbf{Patch size = 512 x 512 pixels ($\sim$3,000 patches per camera per pipeline)}}}                                                                                                                                                                                                                                                                                                                                                                \\ \midrule
    {\color[HTML]{9A0000} \textbf{RT}}                                                                     & \textbf{188}                       & {\color[HTML]{333333} -}           & {\color[HTML]{333333} 163}         & {\color[HTML]{333333} 179}         & {\color[HTML]{333333} 201}         & {\color[HTML]{333333} 213}         & {\color[HTML]{333333} 206}         & {\color[HTML]{333333} 193}         & {\color[HTML]{333333} 164}         & {\color[HTML]{333333} 178}         & {\color[HTML]{333333} 32}          & {\color[HTML]{333333} 14}          & {\color[HTML]{333333} 120}         \\
{\color[HTML]{9A0000} \textbf{AT}}                                                                     & {\color[HTML]{333333} -}           & -                                  & -                                  & -                                  & -                                  & -                                  & -                                  & -                                  & {\color[HTML]{333333} -}           & -                                  & -                                  & -                                  & -                                  \\
{\color[HTML]{9A0000} \textbf{LZ}}                                                                     & {\color[HTML]{333333} 175}         & -                                  & \textbf{186}                       & 177                                & 199                                & 207                                & 190                                & 198                                & {\color[HTML]{333333} 149}         & 192                                & 33                                 & 14                                 & 128                                \\
{\color[HTML]{9A0000} \textbf{DR}}                                                                     & {\color[HTML]{333333} 134}         & -                                  & 124                                & \textbf{146}                       & 166                                & 176                                & 145                                & 137                                & {\color[HTML]{333333} 115}         & 131                                & 25                                 & 14                                 & 103                                \\
{\color[HTML]{9A0000} \textbf{LR}}                                                                     & {\color[HTML]{333333} 175}         & -                                  & 168                                & 192                                & \textbf{214}                       & 225                                & 196                                & 186                                & {\color[HTML]{333333} 153}         & 180                                & 31                                 & 14                                 & 135                                \\
{\color[HTML]{9A0000} \textbf{AF}}                                                                     & {\color[HTML]{333333} 201}         & -                                  & 188                                & 229                                & 254                                & \textbf{269}                       & 230                                & 207                                & {\color[HTML]{333333} 177}         & 201                                & 33                                 & 15                                 & 156                                \\
{\color[HTML]{9A0000} \textbf{LM}}                                                                     & {\color[HTML]{333333} 223}         & -                                  & 203                                & 221                                & 244                                & 268                                & \textbf{332}                       & 239                                & {\color[HTML]{333333} 236}         & 240                                & 25                                 & 15                                 & 153                                \\
{\color[HTML]{9A0000} \textbf{DX}}                                                                     & {\color[HTML]{333333} 177}         & -                                  & 178                                & 166                                & 188                                & 202                                & 198                                & \textbf{225}                       & {\color[HTML]{333333} 165}         & 199                                & 35                                 & 14                                 & 121                                \\ 
{\color[HTML]{9A0000} \textbf{C1}}                                                                     & {\color[HTML]{333333} 93}          & {\color[HTML]{333333} -}           & {\color[HTML]{333333} 85}          & {\color[HTML]{333333} 91}          & {\color[HTML]{333333} 99}          & {\color[HTML]{333333} 107}         & {\color[HTML]{333333} 123}         & {\color[HTML]{333333} 106}         & \textbf{5462}                      & {\color[HTML]{333333} 98}          & {\color[HTML]{333333} 18}          & {\color[HTML]{333333} 13}          & {\color[HTML]{333333} 66}          \\ 
{\color[HTML]{9A0000} \textbf{LT}}                                                                     & {\color[HTML]{333333} 170}         & -                                  & 176                                & 169                                & 191                                & 202                                & 206                                & 208                                & {\color[HTML]{333333} 163}         & \textbf{723}                       & 30                                 & 14                                 & 126                                \\ 
{\color[HTML]{9A0000} \textbf{IN}}                                                                     & {\color[HTML]{333333} 26}          & -                                  & 27                                 & 29                                 & 29                                 & 29                                 & 19                                 & 32                                 & {\color[HTML]{333333} 21}          & 27                                 & \textbf{192}                       & 13                                 & 28                                 \\ 
{\color[HTML]{9A0000} \textbf{UN}}                                                                     & {\color[HTML]{333333} 11}          & -                                  & 11                                 & 14                                 & 13                                 & 13                                 & 12                                 & 12                                 & {\color[HTML]{333333} 11}          & 11                                 & 12                                 & \textbf{22}                        & 15                                 \\ 
{\color[HTML]{9A0000} \textbf{DN}}                                                                     & {\color[HTML]{333333} 124}         & -                                  & 119                                & 136                                & 151                                & 162                                & 141                                & 137                                & {\color[HTML]{333333} 106}         & 131                                & 27                                 & 14                                 & \textbf{138}                       \\
\multicolumn{1}{l}{}                                                                                 & \multicolumn{13}{c}{{\color[HTML]{3531FF} \textbf{Patch size = 256 x 256 pixels ($\sim$13,000 patches per camera per pipeline)}}}                                                                                                                                                                                                                                                                                                                                                              \\ \midrule
{\color[HTML]{9A0000} \textbf{RT}}                                                                     & \textbf{48}                        & {\color[HTML]{333333} -}           & {\color[HTML]{333333} 42}          & {\color[HTML]{333333} 41}          & {\color[HTML]{333333} 46}          & {\color[HTML]{333333} 47}          & {\color[HTML]{333333} 48}          & {\color[HTML]{333333} 48}          & {\color[HTML]{333333} 38}          & {\color[HTML]{333333} 44}          & {\color[HTML]{333333} 14}          & {\color[HTML]{333333} 12}          & {\color[HTML]{333333} 28}          \\ 
{\color[HTML]{9A0000} \textbf{AT}}                                                                     & {\color[HTML]{333333} -}           & -                                  & -                                  & -                                  & -                                  & -                                  & -                                  & -                                  & {\color[HTML]{333333} -}           & -                                  & -                                  & -                                  & -                                  \\ 
{\color[HTML]{9A0000} \textbf{LZ}}                                                                     & {\color[HTML]{333333} 47}          & -                                  & \textbf{48}                        & 42                                 & 46                                 & 47                                 & 46                                 & 50                                 & {\color[HTML]{333333} 37}          & 48                                 & 15                                 & 12                                 & 30                                 \\ 
{\color[HTML]{9A0000} \textbf{DR}}                                                                     & {\color[HTML]{333333} 33}          & -                                  & 31                                 & \textbf{37}                        & 41                                 & 42                                 & 33                                 & 33                                 & {\color[HTML]{333333} 26}          & 32                                 & 13                                 & 12                                 & 26                                 \\ 
{\color[HTML]{9A0000} \textbf{LR}}                                                                     & {\color[HTML]{333333} 44}          & -                                  & 41                                 & 48                                 & \textbf{52}                        & 54                                 & 43                                 & 44                                 & {\color[HTML]{333333} 35}          & 43                                 & 15                                 & 12                                 & 34                                 \\ 
{\color[HTML]{9A0000} \textbf{AF}}                                                                     & {\color[HTML]{333333} 49}          & -                                  & 45                                 & 54                                 & 59                                 & \textbf{62}                        & 50                                 & 49                                 & {\color[HTML]{333333} 39}          & 48                                 & 15                                 & 12                                 & 37                                 \\ 
{\color[HTML]{9A0000} \textbf{LM}}                                                                     & {\color[HTML]{333333} 56}          & -                                  & 52                                 & 49                                 & 55                                 & 56                                 & \textbf{81}                        & 60                                 & {\color[HTML]{333333} 58}          & 61                                 & 13                                 & 12                                 & 35                                 \\ 
{\color[HTML]{9A0000} \textbf{DX}}                                                                     & {\color[HTML]{333333} 47}          & -                                  & 44                                 & 39                                 & 44                                 & 45                                 & 49                                 & \textbf{58}                        & {\color[HTML]{333333} 41}          & 50                                 & 15                                 & 12                                 & 28                                 \\ 
{\color[HTML]{9A0000} \textbf{C1}}                                                                     & {\color[HTML]{333333} 25}          & {\color[HTML]{333333} -}           & {\color[HTML]{333333} 22}          & {\color[HTML]{333333} 23}          & {\color[HTML]{333333} 24}          & {\color[HTML]{333333} 24}          & {\color[HTML]{333333} 31}          & {\color[HTML]{333333} 28}          & \textbf{1517}                      & {\color[HTML]{333333} 26}          & {\color[HTML]{333333} 12}          & {\color[HTML]{333333} 11}          & {\color[HTML]{333333} 19}          \\ 
{\color[HTML]{9A0000} \textbf{LT}}                                                                     & {\color[HTML]{333333} 45}          & -                                  & 44                                 & 39                                 & 43                                 & 44                                 & 51                                 & 53                                 & {\color[HTML]{333333} 40}          & \textbf{195}                       & 14                                 & 12                                 & 28                                 \\ 
{\color[HTML]{9A0000} \textbf{IN}}                                                                     & {\color[HTML]{333333} 13}          & -                                  & 13                                 & 15                                 & 14                                 & 14                                 & 12                                 & 15                                 & {\color[HTML]{333333} 12}          & 13                                 & \textbf{48}                        & 11                                 & 14                                 \\ 
{\color[HTML]{9A0000} \textbf{UN}}                                                                     & {\color[HTML]{333333} 10}          & -                                  & 10                                 & 12                                 & 11                                 & 11                                 & 11                                 & 11                                 & {\color[HTML]{333333} 10}          & 10                                 & 11                                 & \textbf{14}                        & 12                                 \\ 
{\color[HTML]{9A0000} \textbf{DN}}                                                                     & {\color[HTML]{333333} 32}          & -                                  & 30                                 & 35                                 & 37                                 & 39                                 & 32                                 & 33                                 & {\color[HTML]{333333} 25}          & 32                                 & 14                                 & 12                                 & \textbf{33}                        \\
\multicolumn{1}{l}{}    & \multicolumn{13}{c}{{\color[HTML]{3531FF} \textbf{Patch size = 128 x 128 pixels ($\sim$55,000 patches per camera per pipeline)}}}                                                                                                                                                                                                                                                                                                                                                              \\ \midrule
{\color[HTML]{9A0000} \textbf{RT}}                                                                     & \textbf{12}                        & {\color[HTML]{333333} -}           & {\color[HTML]{333333} 12}          & {\color[HTML]{333333} 12}          & {\color[HTML]{333333} 11}          & {\color[HTML]{333333} 12}          & {\color[HTML]{333333} 11}          & {\color[HTML]{333333} 12}          & {\color[HTML]{333333} 11}          & {\color[HTML]{333333} 11}          & {\color[HTML]{333333} 10}          & {\color[HTML]{333333} 11}          & {\color[HTML]{333333} 12}          \\ 
{\color[HTML]{9A0000} \textbf{AT}}                                                                     & {\color[HTML]{333333} -}           & -                                  & -                                  & -                                  & -                                  & -                                  & -                                  & -                                  & {\color[HTML]{333333} -}           & -                                  & -                                  & -                                  & -                                  \\ 
{\color[HTML]{9A0000} \textbf{LZ}}                                                                     & {\color[HTML]{333333} 17}          & -                                  & \textbf{21}                        & 16                                 & 16                                 & 16                                 & 14                                 & 17                                 & {\color[HTML]{333333} 14}          & 15                                 & 12                                 & 13                                 & 14                                 \\ 
{\color[HTML]{9A0000} \textbf{DR}}                                                                     & {\color[HTML]{333333} 13}          & -                                  & 14                                 & \textbf{16}                        & 15                                 & 15                                 & 12                                 & 14                                 & {\color[HTML]{333333} 12}          & 13                                 & 11                                 & 13                                 & 14                                 \\ 
{\color[HTML]{9A0000} \textbf{LR}}                                                                     & {\color[HTML]{333333} 16}          & -                                  & 16                                 & 19                                 & \textbf{20}                        & 20                                 & 14                                 & 16                                 & {\color[HTML]{333333} 13}          & 14                                 & 12                                 & 14                                 & 15                                 \\ 
{\color[HTML]{9A0000} \textbf{AF}}                                                                     & {\color[HTML]{333333} 15}          & -                                  & 16                                 & 18                                 & 18                                 & \textbf{20}                        & 13                                 & 15                                 & {\color[HTML]{333333} 13}          & 14                                 & 12                                 & 14                                 & 15                                 \\ 
{\color[HTML]{9A0000} \textbf{LM}}                                                                     & {\color[HTML]{333333} 11}          & -                                  & 11                                 & 12                                 & 11                                 & 11                                 & \textbf{12}                        & 12                                 & {\color[HTML]{333333} 12}          & 12                                 & 10                                 & 11                                 & 12                                 \\ 
{\color[HTML]{9A0000} \textbf{DX}}                                                                     & {\color[HTML]{333333} 15}          & -                                  & 16                                 & 15                                 & 14                                 & 15                                 & 17                                 & \textbf{25}                        & {\color[HTML]{333333} 16}          & 17                                 & 12                                 & 12                                 & 13                                 \\ 
{\color[HTML]{9A0000} \textbf{C1}}                                                                     & {\color[HTML]{333333} 10}          & {\color[HTML]{333333} -}           & {\color[HTML]{333333} 11}          & {\color[HTML]{333333} 11}          & {\color[HTML]{333333} 11}          & {\color[HTML]{333333} 11}          & {\color[HTML]{333333} 11}          & {\color[HTML]{333333} 11}          & \textbf{397}                       & {\color[HTML]{333333} 11}          & {\color[HTML]{333333} 10}          & {\color[HTML]{333333} 10}          & {\color[HTML]{333333} 11}          \\ 
{\color[HTML]{9A0000} \textbf{LT}}                                                                     & {\color[HTML]{333333} 11}          & -                                  & 12                                 & 12                                 & 11                                 & 12                                 & 12                                 & 13                                 & {\color[HTML]{333333} 11}          & \textbf{24}                        & 10                                 & 11                                 & 11                                 \\ 
{\color[HTML]{9A0000} \textbf{IN}}                                                                     & {\color[HTML]{333333} 10}          & -                                  & 10                                 & 11                                 & 10                                 & 11                                 & 10                                 & 11                                 & {\color[HTML]{333333} 10}          & 10                                 & \textbf{12}                        & 10                                 & 11                                 \\ 
{\color[HTML]{9A0000} \textbf{UN}}                                                                     & {\color[HTML]{333333} 13}          & -                                  & 13                                 & 14                                 & 14                                 & 14                                 & 12                                 & 12                                 & {\color[HTML]{333333} 12}          & 12                                 & 11                                 & \textbf{15}                        & 13                                 \\ 
{\color[HTML]{9A0000} \textbf{DN}}                                                                     & {\color[HTML]{333333} 11}          & -                                  & 11                                 & 12                                 & 11                                 & 12                                 & 11                                 & 11                                 & {\color[HTML]{333333} 11}          & 11                                 & 10                                 & 11                                 & \textbf{12}                        \\
 \bottomrule
    \end{tabular}
    % }
    \end{table*}
\begin{figure*}[htb!]
    \centering
    \includegraphics[width=\textwidth] {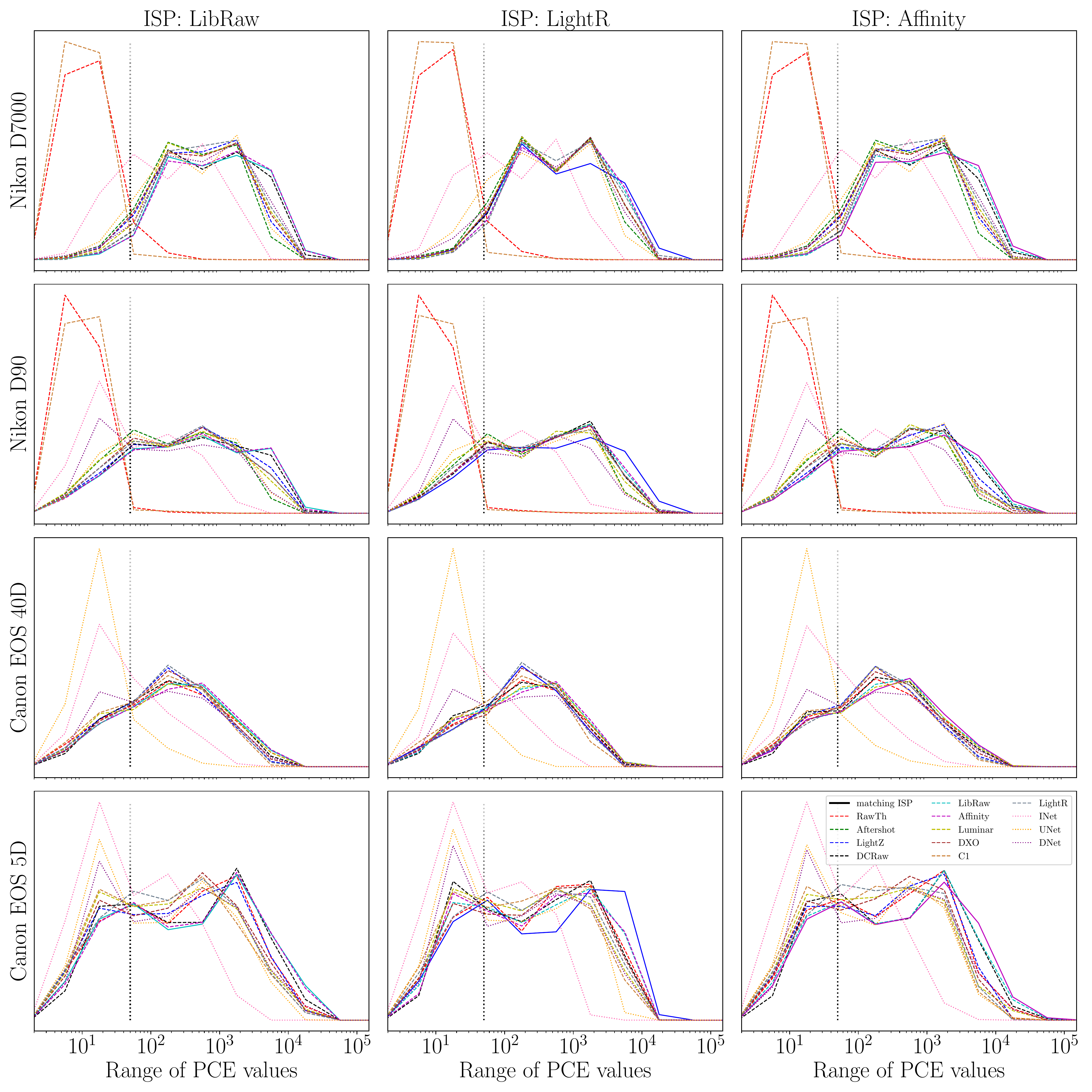}
    \caption{Distribution of PCE scores for cross-ISP matching (512~px patches); the dashed vertical line shows a common acceptance threshold of 50.}
    \label{fig:ISP_PCEDist_Libraw_Supp}
\end{figure*}

\begin{figure*}[htb!]
    \centering
    \includegraphics[width=\textwidth]{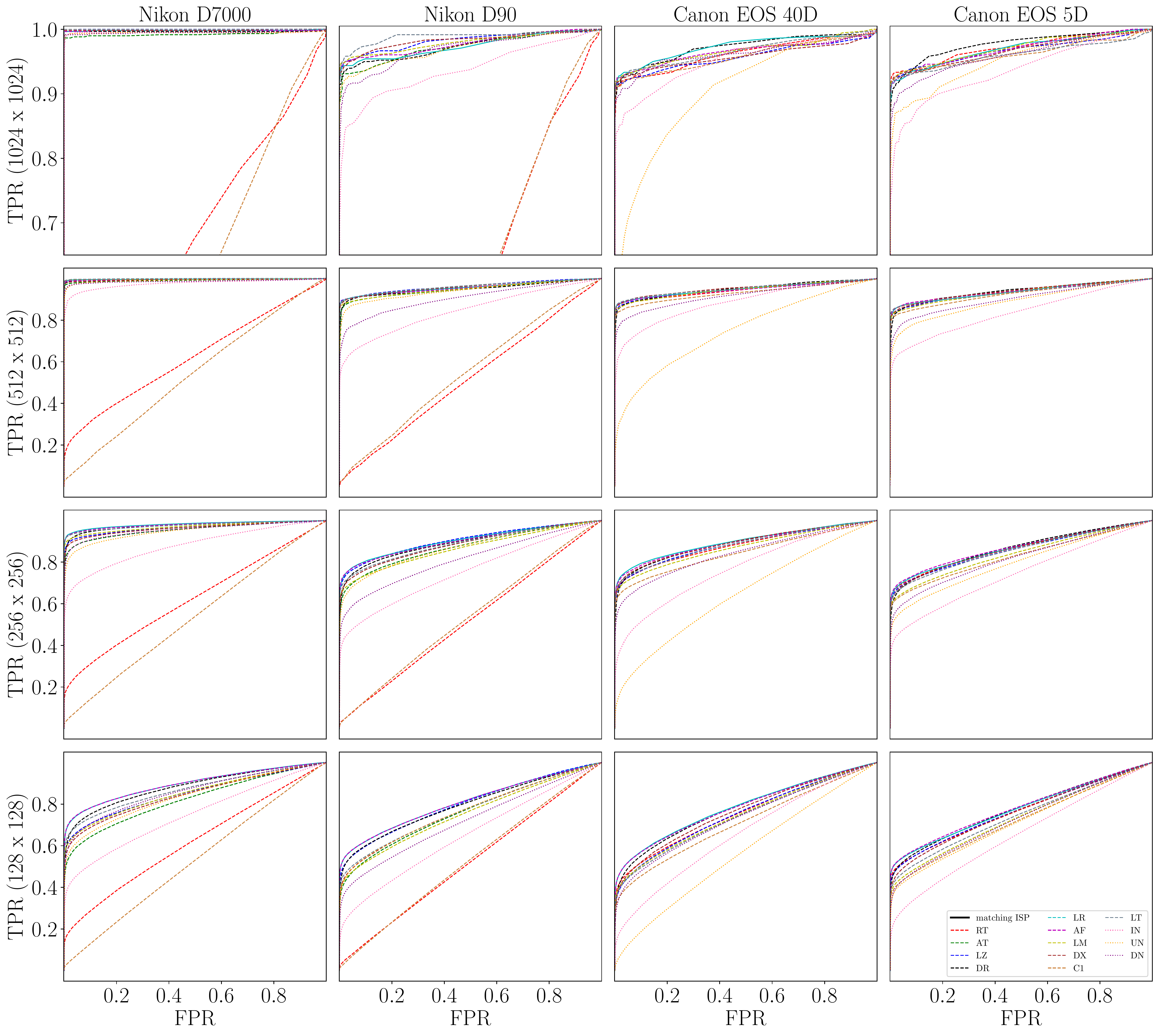}
    \caption{Receiver operating characteristic (ROC) curves with LibRaw as PRNU estimation ISP and all pipelines as test.}
    \label{fig:ISP_ROC_PatchSizes_Supp}
\end{figure*}

\end{document}